\documentclass[letterpaper,twocolumn,10pt]{article}
\usepackage{usenix2019_v3}

\usepackage{tikz}
\usepackage{amsmath}
\usetikzlibrary{decorations.pathreplacing}

\usepackage{filecontents}
\usepackage{tikz}
\usetikzlibrary{
    positioning,
    fit,
    arrows.meta,
    shapes.geometric,
    calc
}
\usepackage{amsmath}
\usepackage{tikz}
\usepackage{amsfonts}
\usepackage{makecell}
\usepackage{hyperref} 
\usepackage{accents}
\usepackage{caption}
\usepackage{subcaption}
\usepackage{graphicx}
\usepackage{xspace}
\usepackage{url}

\newcommand{\etal}{{et~al.}} 
\newcommand{\ie}{{i.e.,~}}
\newcommand{\eg}{{e.g.,~}}
\DeclareMathSymbol{\mathbbE}{\mathord}{AMSb}{"45}

\newcommand{\TT}[1]{``\textit{#1}''}

\newcommand{\todobox}[3]{%
	\colorbox{#1}{\textcolor{white}{\sffamily\bfseries\scriptsize #2}}%
	~\textcolor{red}{#3} %
	\textcolor{#1}{$\triangleleft$}%
}
\newcommand{\todo}[1]{\todobox{red}{TODO}{#1}}

\usepackage{tabularx,booktabs} 

\usepackage{wasysym} 

\newcommand{\user}{\mathit{u}}
\newcommand{\agent}{red\mathsf{A}}
\newcommand{\umachine}{\mathit{uM}}
\newcommand{\target}{\mathit{T}}
\newcommand{\adv}{\mathcal{A}}

\newcommand{\agentname}{\textit{agentic-red-team}\xspace}
\newcommand{\agentnames}{\textit{agentic-red-teams}\xspace}
\newcommand{\nametrojan}{agent-phishing\xspace}

\newcommand{\metatron}{\texttt{METATRON}\xspace}
\newcommand{\nebula}{\texttt{nebula}\xspace}
\newcommand{\xalgorix}{\texttt{xalgorix}\xspace}
\newcommand{\cai}{\texttt{CAI}\xspace}

\newcommand{\AIRecong}{\texttt{AIRecon}\xspace}
\newcommand{\PentestGPT}{\texttt{PentestGPT}\xspace}
\newcommand{\pentagi}{\texttt{PentAGI}\xspace}
\newcommand{\RedAmon}{\texttt{RedAmon}\xspace}
\newcommand{\strix}{\texttt{STRIX}\xspace}
\newcommand{\artemis}{\texttt{Artemis}\xspace}

\begin{document}

\date{}

\title{Red-Teaming the Agentic Red-Team}

\author{
{\rm Dario Pasquini\thanks{Corresponds to:  \url{dp@cracken.ai}}}\\
Cracken
\and
{\rm Michał Bazyli}\\
Cracken
\and
{\rm Taras Fedynyshyn}\thanks{
Information Protection Department,
Lviv Polytechnic National University, Lviv, 79013, Ukraine}\\
Cracken
\and
{\rm Artem Sorokin}\\
Cracken
} 

\maketitle
\begin{abstract}
The use of agentic systems to perform offensive security operations has moved from a theoretical possibility to a commoditized capability.
  However, while the community has focused on creating more and more capable agents, less attention has been allocated to assessing
  the security of those systems.
  
In this work, we present the first in-depth security analysis of the most widely used agentic systems for offensive security operations. We show that most of these tools share common design flaws that enable an active adversary to exfiltrate API keys, establish persistent footholds, and fully compromise the operator's machine, even when the agent operates inside a sandboxed container.
  
To support our analysis, we introduce a full cyber kill chain for such agentic systems, capturing the progression from initial LLM manipulation to lateral movement, persistence, guardrail bypass, and sandbox escape.

Building on our security analysis, we derive a robust architecture for agentic offensive-security tools and propose actionable, broadly applicable design principles that mitigate the disclosed attack paths at the architectural level.
 \end{abstract}


\section{Introduction}
Since the inception of machine learning, both researchers and industry practitioners have sought to leverage models to automate security operations. Early deep learning-based solutions were largely confined to narrow subroutines and specific tasks such as password guessing~\cite{melicher2016fast, pasquini2024universal, pasquini2021improving} and fuzzing~\cite{bottinger2018deep, zong2020fuzzguard}, among many others~\cite{shaukat2020survey}.

The introduction of LLMs and derived agentic systems has enabled a fundamental leap, unlocking the possibility of end-to-end, full-cycle security operations with minimal or no human involvement. By far, the most prominent application in both research~\cite{deng2024pentestgpt, Happe_2023, fang2024llmagentsautonomouslyexploit, fang2024llmagentsautonomouslyhack, autoattacker, zhu2026teams, wang2024sands,shao2024empiricalevaluationllmssolving, liu2026synthesizing, singer2025incalmo, shi2026cybergym} and industry~\cite{xbow2026, cracken2026, wiz2026redagent, novee2026, aliasrobotics2026alias1, aikido2026aipentest, lin2025comparing, wang2026exploitgym} is the development of offensive security systems capable of performing tasks such as automated penetration testing, red teaming, and full-chain adversary emulation. Moreover, as these systems are becoming ubiquitous, they are also transcending their role as tools for exploratory security analysis and are increasingly being deployed in real-world operational contexts, either in support of cyber forces~\cite{schere2026seekr, freedberg2026armyaitx, Kovacs2026LockedShields} or, more critically, for malicious use in cyberattacks~\cite{google2026aivulninitialaccess, anthropic2025espionage}.

As these tools rapidly become core components of security operations and are increasingly relied upon by operators~\cite{NEURIPS2025_faed4276}, we must ask whether they are sufficiently secure to withstand real-world exposure and, thus, the scrutiny of adversaries.

In this work, we present the first in-depth security analysis of this emerging class of tools. We demonstrate that they introduce new attack vectors against the organizations and users who deploy them.\footnote{A common pattern shared with other emerging agentic systems~\cite{aioops, nassi2025invitation, ICLR2025_42f92b78, lee2025takedown}.}
\begin{table}[t]
  \centering
  \caption{Attacker capabilities across audited agentic offensive-security systems. Legend: \CIRCLE\ deterministic, \LEFTcircle\ depending on external factors, but likely, \Circle\ no reliable exploit found, $^{\dagger}$~no OS-level sandbox (host access immediate).}
  \label{tab:vuln-matrix}
  
  \resizebox{1\columnwidth}{!}{
  \begin{tabular}{@{}lccccc@{}}
    \toprule
    \textbf{Agent} & \textbf{RCE on} & \textbf{Secrets} & \textbf{Persis-} & \textbf{Unbounded} & \textbf{Host} \\
                   & \textbf{worker} & \textbf{exfil.} & \textbf{tence} & \textbf{weaponization} & \textbf{compromise} \\
    \midrule
    CAI                  & \CIRCLE & \CIRCLE\ All         & \CIRCLE\ hard & \CIRCLE & \CIRCLE\ RCE \\
    RedAmon              & \CIRCLE & \CIRCLE\ LLM keys ++ & \CIRCLE\ hard & \CIRCLE & \CIRCLE\ Network \\
    PentestAgent         & \CIRCLE & \CIRCLE\ All         & \CIRCLE\ hard & \CIRCLE & \CIRCLE\ RCE \\
    DarkMoon             & \CIRCLE & \CIRCLE\ All         & \CIRCLE\ hard & \CIRCLE & \CIRCLE\ RCE \\
    PentAGI              & \CIRCLE & \CIRCLE\ All         & \CIRCLE\ hard & \CIRCLE & \CIRCLE\ RCE \\
    AIRecon              & \CIRCLE & \CIRCLE\ All         & \CIRCLE\ hard & \CIRCLE & \CIRCLE\ RCE \\
    PentestGPT           & \CIRCLE & \CIRCLE\ LLM keys    & \CIRCLE\ hard & \CIRCLE & \LEFTcircle\ RCE \\
    METATRON$^{\dagger}$ & \CIRCLE & \CIRCLE\ All         & \CIRCLE\ hard & \CIRCLE & \CIRCLE\ RCE \\
    nebula$^{\dagger}$   & \CIRCLE & \CIRCLE\ All         & \CIRCLE\ hard & \CIRCLE & \CIRCLE\ RCE \\
    xalgorix$^{\dagger}$ & \CIRCLE & \CIRCLE\ All         & \CIRCLE\ hard & \CIRCLE & \CIRCLE\ RCE \\
    Artemis              & \CIRCLE & \CIRCLE\ LLM keys    & \CIRCLE\ soft & \CIRCLE & \Circle \\
    STRIX                & \CIRCLE & \Circle     & \Circle & \CIRCLE & \Circle \\
    \bottomrule
  \end{tabular}
  }
\end{table}
 Specifically, we show that an adversary who controls the target of an offensive-security operation can leverage this position to manipulate the agent through techniques adjacent to prompt injection, ultimately achieving arbitrary code execution on the user's own infrastructure regardless of sandboxing. 
\begin{table*}[t]

\centering
\scriptsize
\setlength{\tabcolsep}{5pt}
\renewcommand{\arraystretch}{1.15}

\resizebox{1\textwidth}{!}{
\begin{tabular}{lccccccccl}

\hline

\textbf{Agent} &
\textbf{Sandbox} &
\textbf{Guardrail} &
\textbf{Human} &
\textbf{Memory} &
\textbf{Memory /} &
\textbf{Multi-} &
\textbf{Skills} &
\textbf{UI} &
\textbf{Agentic Harness} \\

& & & \textbf{gate} & & \textbf{sessions} & \textbf{agent} & & & \\

\hline

AIRecon         & $\textit{Docker}$     & $\checkmark^{*}$ & $-$ & $\checkmark$ & $\checkmark$ & $\checkmark$ & $\checkmark$ & TUI             & Custom (Python / Ollama) \\

CAI              & $\textit{Docker}^{*}$ & $\checkmark^{*}$ & $-$ & $\checkmark$ & $\checkmark$ & $\checkmark$ & $\checkmark$ & CLI (REPL)      & Custom (LiteLLM / OpenAI SDK) \\

PentAGI          & $\textit{Docker}$     & $-$         & $-$ & $\checkmark$ & $\checkmark$ & $\checkmark$ & $-$     & Web             & Custom (Go) \\

RedAmon          & $\textit{Docker}$     & $\checkmark$     & $\checkmark^{*}$ & $\checkmark$ & $\checkmark$ & $\checkmark$ & $\checkmark$ & Web & LangGraph \\

STRIX            & $\textit{Docker}$     & $-$         & $-$ & $\checkmark$ & $-$     & $\checkmark$ & $\checkmark$ & CLI / TUI       & Custom (LiteLLM) \\

Artemis          & $\textit{Docker}^{*}$ & $-$         & $-$ & $\checkmark$ & $\checkmark$ & $\checkmark$ & $-$     & CLI             & Codex \\

METATRON         & $-$         & $\checkmark$     & $-$ & $\checkmark$ & $\checkmark$ & $-$     & $-$     & CLI             & Custom (Python)  \\

PentestAgent     & $\textit{Docker}^{*}$ & $\checkmark$     & $-$ & $\checkmark$ & $\checkmark$ & $\checkmark$ & $\checkmark$ & TUI             & Custom (Python / LiteLLM) \\

PentestGPT       & $\textit{Docker}$     & $-$         & $-$ & $\checkmark$ & $\checkmark$ & $-$     & $-$     & TUI             & Claude Code \\

nebula           & $-$         & $-$         & $-$ & $\checkmark$ & $\checkmark$ & $-$     & $-$     & Desktop (PyQt6) & LangChain \\

xalgorix         & $-$         & $\checkmark$     & $-$ & $\checkmark$ & $-$     & $\checkmark$ & $\checkmark$ & Web / CLI       & Custom (Go) \\

DarkMoon         & $\textit{Docker}$     & $\checkmark$     & $-$ & $-$     & $-$     & $\checkmark$ & $\checkmark$ & CLI             & OpenCode \\

\hline

\end{tabular}
}

\caption{List of \agentnames under consideration and their relevant features. $\checkmark$ indicates the feature is present, $-$ indicates absence, and ``$^{*}$'' indicates the feature is optional or configurable.}

\label{tab:agent-comparison}
\end{table*}

In our security analysis spanning 12 harnesses and 7 LLMs, we introduce several attacks and exploits. Primarily, we present a prompt-injection-free manipulation attack design for offensive security agents that achieves near-deterministic remote code execution on the agent infrastructure, even when deployed with frontier LLMs such as \texttt{Claude Opus 4.8}, \texttt{GPT-5.5}, and \texttt{Gemini 3.1 Pro}. The attack does not rely on explicit prompt-injection payloads~\cite{greshake2023not, pasquini2024neural} and instead leverages contextual deception and reward hacking as the main drivers~\cite{mantis}. Drawing inspiration from malware antivirus-evasion techniques, we show how this technique can encode malicious functionality within payloads capable of bypassing both frontier model safeguards and their direct scrutiny with high confidence.

Beyond LLM-targeting attacks, we identify recurring insecure design patterns in the tested tools (summarized in Table~\ref{tab:vuln-matrix}). Once an attacker achieves code execution within the execution environment (typically a sandbox or container) by exploiting the agent, weak segmentation and violations of the principle of least privilege enable escalation across the system, allowing secrets exfiltration, agent weaponization against arbitrary targets, and, in most cases (10/12), sandbox escape leading to host compromise.

Motivated by these recurrent flaws, we define a tailored kill chain to model such escalation processes and paths. The objective of this kill chain is to enable more principled and systematic red teaming of such systems, providing an abstraction that generalizes both attack vectors and attacker objectives.


Finally, drawing on our empirical results and threat modeling, we propose a robust and general architecture designed to prevent escalation within the kill chain. The core design principle here is: rather than assuming that LLMs can be reliably hardened against manipulation techniques such as prompt injection, we adopt a stronger adversarial model in which we assume that the LLM will behave arbitrarily and maliciously. We then \textbf{design the system architecture around the central invariant of minimizing and containing the blast radius of an untrusted LLM / worker environment. This approach yields security guarantees that preserve soundness even if prompt injection is never fundamentally solved}~\cite{abdelnabi2026aiagentsfallprompt, christodorescu2026agentsecuritysystemsproblem}.

Our contributions can be then summarized as follows:
\begin{itemize}
\itemsep0em
	\item We perform the first in-depth analysis of the security properties of agentic offensive-security tools, both at the AI and system level. We abstract our findings into a full-cycle cyber kill chain tailored for such tools, providing a systematic roadmap for threat modeling and red teaming them.

	\item We introduce a prompt-injection-free manipulation technique that achieves near-deterministic effectiveness against agentic offensive-security tools.

	\item Finally, we design a principled secure architecture for such systems, based on the invariant of containment and least privilege, rather than on optimistic assumptions that AI-targeted attacks such as prompt injection can be prevented.
\end{itemize}

In addition, based on the LLM manipulation techniques built within the scope of this security analysis, we implement and open-source \textbf{BlackSea}---an active host-based deception system aimed at defending against LLM-driven cyberattacks: \url{https://github.com/cracken-ai/blacksea}.


\section{Preliminary}
\label{sec:preliminary}
This section provides the necessary background and outlines the setup of our study.

\subsection{Agentic Offensive Security}
\label{sec:agentic_offensive_security}
We define an agentic offensive-security system as any agentic system designed to perform or support offensive security tasks in an autonomous manner through one or more LLMs~\cite{deng2024pentestgpt, Happe_2023, fang2024llmagentsautonomouslyexploit, fang2024llmagentsautonomouslyhack, autoattacker, zhu2026teams, wang2024sands,shao2024empiricalevaluationllmssolving, liu2026synthesizing, singer2025incalmo}.

In this study, we primarily focus on fully autonomous systems, as they represent the vast majority of available ones. We focus on agents that are explicitly marketed to support black-box offensive operations, which constitute the most common class.\footnote{That is, we do not consider tools that perform \textbf{only} code review or general white-box code analysis.} For convenience, without loss of generality, we hereafter refer to this class of systems as \agentnames.

In our study, we consider 12 open-source \agentnames, representing the most popular systems available at the time of writing. The complete list of agents and their main features is provided in Table~\ref{tab:agent-comparison}, while repository and popularity details are reported in Table~\ref{tab:agents_git} in Appendix~\ref{app:add_material}.

All considered agents are designed to be fully autonomous and are primarily intended for black-box security analysis, while also presenting tooling for white-box operations.


\subsubsection{Architecture Abstraction}
\begin{figure}

\centering

\resizebox{.9\columnwidth}{!}{

\begin{tikzpicture}[
    font=\small,
    node distance=1.0cm and 1.4cm,
    box/.style={draw, rounded corners, align=center, minimum width=3.2cm, minimum height=1cm},
    micro/.style={draw, rounded corners, align=center, minimum width=2.2cm, minimum height=0.7cm, font=\scriptsize},
    arrow/.style={-{Stealth[length=2mm]}, thick},
    darrow/.style={{Stealth[length=2mm]}-{Stealth[length=2mm]}, thick}
]

\node[micro] (llm) {Agentic harness};

\node[micro, dotted, below=of llm, yshift=0.9cm] (db) {Skills / DB / Memory / Orch. tools};

\node[below=of db, yshift=0.9cm, draw=black] (guard) {\scriptsize{guardrails}};

\node[draw, rounded corners,
      fit=(llm) (guard),
      inner sep=0cm, inner xsep=0.8cm] (orchbox) {};

\node at (orchbox.north) [above] {Orchestrator:};

\node[box, below=of orchbox, minimum width=2.8cm, minimum height=1cm] (worker)
{
Worker\\
{\scriptsize (bash execution runtime)}
};

\node[draw, dashed, rounded corners,
      fit=(orchbox) (worker),
      inner xsep=0.7cm, yshift=0.15cm, inner ysep=0.3cm,] (host) {};
\node at (host.north) [above] {$\umachine$};


\draw[arrow] (orchbox.south) -- node[right, font=\scriptsize, align=center]
{tool call\\(bash cmd)} (worker);

\node[box, right=2cm of worker, minimum width=1cm, draw=red, fill=white] (target) {$\target$\\(attacker)};

\draw[darrow, dashed, red] (worker) -- node[above, font=\scriptsize]
{network} (target);

\node[box, right=of llm, xshift=-.3cm, yshift=-0.5cm, fill=white, minimum width=1cm] (frontend) {UI};

\node[box, right=of frontend, xshift=-.9cm, fill=white, minimum width=1cm, draw=green] (user) {$\user$\\(victim)};
\draw[arrow] (frontend) -- (orchbox);
\draw[arrow] (user) -- (frontend);

\end{tikzpicture}
}

\caption{Abstraction of the architecture of a \agentname, and the parties involved in the threat model.}
\label{fig:basic_arch}

\end{figure}
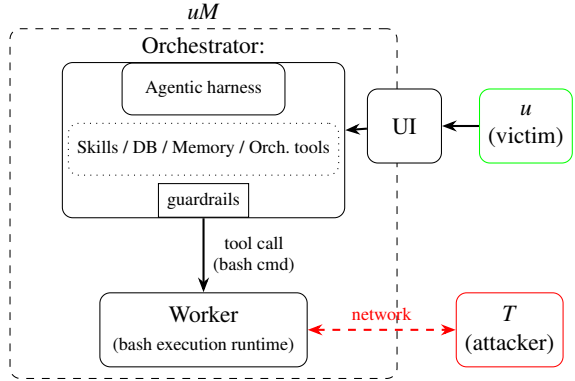
The architecture of \agentnames{} drastically varies across implementations; however, these systems can generally be abstracted as comprising three main components, as depicted in Figure~\ref{fig:basic_arch}.

\paragraph{1) Orchestrator:}
The orchestrator is the software/infrastructure  layer that implements and coordinates the agentic system. It defines and manages the components that support the agentic harness (e.g., session memory, episodic memory, skills, and MCP access), as well as the logic governing their interactions. This is where the main agentic loop runs and where tool calls are issued for execution by the second component, the \TT{worker}.

In addition, the orchestrator may itself execute certain tools, typically those related to memory operations or system management (e.g., spawning workers). It also enforces policies and guardrails when present, such as preventing the LLM from operating on out-of-scope targets or executing destructive actions.

At the infrastructural level, the orchestrator provides persistence mechanisms, typically through a database or a core file system, which store session state, intermediate artifacts, and other persistent data required for the agent's operation.

The orchestrator can either run in a container or directly as a process on the operator's machine~$\umachine$. In containerized deployments, it is also responsible for transmitting relevant data (typically reports and other artifacts) to the operator's machine~$\umachine$, where it can be accessed by the user $\user$.

\paragraph{2) Worker:}
The worker (the execution environment) is the component responsible for executing tools issued by the orchestrator, typically in the form of raw shell commands. In most cases, it consists of a \textit{Kali Linux} environment equipped with a suite of security tools that the agent can freely orchestrate. Upon receiving a tool call from the orchestrator, this is translated into an executable operation for the worker (e.g., a complete Bash command string), which is then executed accordingly. Some \agentnames support the use of multiple workers simultaneously; typically, different containers with different capabilities.

Workers are generally containerized systems (\eg docker containers) running within the operator's machine~$\umachine$. Typically, they are the only components of the system that directly interact with the target~$\target$ and the external world (i.e., external to~$\umachine$).

\paragraph{3) Front-end:}
This component handles the interaction between the user and the orchestrator. Such systems typically operate in one of two modes: a command-line interface  or a web-based interface. The front-end is mostly irrelevant to our analysis and is therefore largely omitted in the remainder of this work. It is only discussed when it forms part of an exploit chain.

Hereafter, we apply such abstraction to model the agentic systems we analyze.

\subsection{Threat model}
\label{sec:threat_model}
\textbf{Premise:} We frame our security analysis around an attacker who controls the target of the offensive-security operation run by the \agentname (see Figure~\ref{fig:basic_arch}), with the goal of compromising its operator. While this threat model has practical relevance, for instance in modeling hack-backs within an active cyber conflict, we adopt it primarily as a principled lens to study the security of these systems, and our conclusions generalize beyond this specific setting.\\

Formally: we model a user~$\user$ running an \agentname~$\agent$ on a machine~$\umachine$ (the operator's machine) to perform an offensive security operation on a remote target~$\target$ (e.g., a web application or a remote machine). The target~$\target$ is controlled by the adversary~$\adv$, whose objective is to compromise the user~$\user$ by indirectly exploiting the agent~$\agent$ as an attack vector.

\paragraph{Attacker objectives:}
In this work, we primarily model an attacker who seeks \textbf{full operator's machine compromise}; that is, the attacker escalates privileges and, where necessary, escapes the agent's sandbox to achieve code execution directly on the operator's machine $\umachine$ (also referred to as \textit{host}).

Additionally, we also model other intermediate objectives: \textit{persistent and unbounded agent weaponization} and \textit{secrets exfiltration}, which will be discussed in \S~\ref{sec:kill_chain}. 

\paragraph{Attacker ($\adv$) capabilities}
We model a weak adversary with no prior knowledge about the user~$\user$ or the deployed agentic system. The attacker's capabilities are restricted to arbitrary manipulation of the target system~$\target$, which is under their control. No additional capabilities are assumed or granted.
 
 \paragraph{\agentname ($\agent$) deployment:}
 In our study, we deploy the tools listed in Table~\ref{tab:agent-comparison} under their default configurations, while assuming that the strongest available security mechanisms are enabled whenever they are optional. Specifically, we adhere to the following rules: \textbf{(1)} if a system supports optional, non-default sandboxing, we assume that sandboxing is enabled. We do not run the system without a sandbox unless the documentation or code contains no mention of sandboxing capabilities.
\textbf{(2)}~if a system provides optional, non-default guardrails, we assume that these guardrails are active. \textbf{(3)}~we assume that worker containers (see \S~\ref{sec:agentic_offensive_security}) are reset and their filesystems cleaned after each operation, even if this practice is not explicitly documented by the developers (see persistent weaponization in \S~\ref{sec:weaponization}). \textbf{(4)}~if services are shipped with default credentials (e.g., database passwords), we assume that these credentials have been rotated.

None of these tools involve any form of human gating or decision delegation; therefore, we are not required to make any assumptions about human behavior within the threat model.

\subsubsection{Why the security of offensive security tools matters}
Before turning to our findings, we want to make explicit the motivations behind our study. We argue that there are at least two core reasons why the security of agentic offensive-security tools is a relevant and timely problem to study:

\paragraph{Agentic Offensive-Security Systems will be deployed in the field:}
Automated penetration testing and red-teaming systems have rapidly emerged as a commoditized capability within months of the introduction of sufficiently capable foundation models~\cite{xbow2026, cracken2026, novee2026, aliasrobotics2026alias1, aikido2026aipentest, lin2025comparing, wang2026exploitgym, wiz2026redagent}. Their adoption is accelerating at a rapid pace, with both capability and deployment frequency increasing continuously. As a result, an increasing number of organizations are incorporating these systems into production security workflows.

Cyber forces are also following this trend, showing growing interest in integrating these technologies into field operations~\cite{schere2026seekr, freedberg2026armyaitx, Kovacs2026LockedShields}. If this is not already the case today, it is highly likely that agentic tools will soon become embedded components in cyber conflict scenarios, either as core operational assets or as force multipliers supporting field operators. 

As we move toward such a setting, ensuring that these systems remain secure under adversarial exposure becomes a fundamental requirement; that is, in such environments, adversarial targeting is not a hypothetical possibility, but an expected condition of deployment.

 \paragraph{Beyond the Adversarial Setting:}
 The secure design of offensive-security agents is necessary even under the optimistic assumption that no active adversary is present. Such systems operate with high-impact capabilities and privileges (see \S~\ref{sec:sandbox_escape_via_cap}), including network access, arbitrary code execution, and interaction with sensitive infrastructure. Consequently, the same failure that an adversary could induce on purpose can also arise on its own: a hallucination, a planning error, or reward misalignment can drive the agent into an unsafe state and directly produce destructive, irreversible outcomes, as repeatedly demonstrated by incidents involving production autonomous systems \cite{fortune2025replit, IST2026AILossOfControl}.

 For this reason, adversarial robustness should be read as a worst-case model of agent misbehavior rather than a niche concern. In the threat model considered in this work, the adversary explicitly attempts to drive the agent toward unsafe states and unauthorized actions; this is precisely the behavior that hallucinations and misalignment can trigger accidentally. A system that stays contained and limits damage under deliberate adversarial manipulation is therefore, by construction, also robust against naturally occurring failures such as unintended autonomous behavior. In other words, the conclusions we derive from our study are not specific to the adversarial case: they are a prerequisite for the safe operation of any offensive-security agent, adversary or not.

\section{The Kill-Chain}
\label{sec:kill_chain}
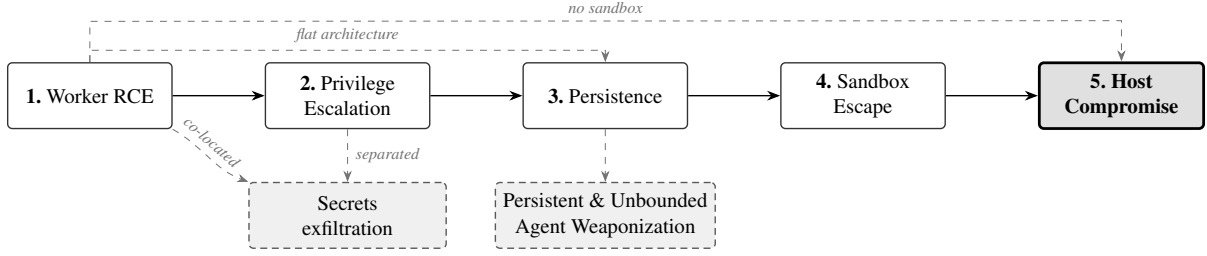
\begin{figure*}[t]
	
	\centering

\resizebox{.9\textwidth}{!}{
\begin{tikzpicture}[
      node distance=0.9cm and 1.4cm,
      stage/.style={           
          rectangle, rounded corners=2pt,
          draw=black!80, fill=white, line width=0.6pt,
          minimum width=2.5cm, minimum height=1.0cm,
          align=center, font=\small
      },
      terminal/.style={        
          rectangle, rounded corners=2pt,
          draw=black, fill=black!12, line width=1.2pt,
          minimum width=2.5cm, minimum height=1.0cm,
          align=center, font=\small\bfseries
      },
      objective/.style={       
          rectangle, rounded corners=2pt,
          draw=black!65, fill=black!6, line width=0.6pt,
          minimum width=3.0cm, minimum height=1.0cm,
          align=center, font=\small, densely dashed
      },
      arr/.style={-Stealth, semithick},
      darr/.style={-Stealth, dashed, black!55, thin},
  ]

  \node[stage]    (rce)  {\textbf{1.} Worker RCE};
  \node[stage,    right=of rce]  (priv) {\textbf{2.} Privilege\\Escalation};
  \node[stage,    right=of priv] (per)  {\textbf{3.} Persistence};
  \node[stage,    right=of per]  (sand) {\textbf{4.} Sandbox\\Escape};
  \node[terminal, right=of sand] (host) {\textbf{5.} Host\\Compromise};

  \draw[arr] (rce)  -- (priv);
  \draw[arr] (priv) -- (per);
  \draw[arr] (per)  -- (sand);
  \draw[arr] (sand) -- (host);

  \draw[darr]
      (rce.north) -- ++(0, 0.62)
      -- node[above, font=\scriptsize, midway] {\textit{no sandbox}}
         ($(host.north) + (0, 0.62)$)
      -- (host.north);
      
    \draw[darr]
      (rce.north) -- ++(0, 0.18)
      -- node[above, font=\scriptsize, midway] {\textit{flat architecture}}
         ($(per.north) + (0, 0.18)$)
      -- (per.north);

  \node[objective, below=.8cm of priv] (cred) {Secrets\\exfiltration};
  \node[objective, below=.8cm of per]  (weap) {Persistent \& Unbounded \\ Agent Weaponization};

  \draw[darr] (per.south) -- (weap.north);

  \draw[darr] (rce.south east)
      to[out=-30, in=160]
      node[above, font=\scriptsize, sloped, pos=0.4] {\textit{co-located}}
      (cred.north west);

  \draw[darr] (priv.south)
      -- node[right, font=\scriptsize, pos=0.5] {\textit{separated}}
      (cred.north);

  \end{tikzpicture}
  }
  
  \caption{Kill chain for \agentnames (first row), and intermediate adversarial objectives (second row).  Dotted lines represent skipping connections conditionally enabled by the target architecture.}
  \label{fig:kill_chain}
  
 \end{figure*}
In our security analysis, we notice a recurring set of vulnerabilities and misconfigurations across  \agentnames. These shared weaknesses enable attackers to compromise such systems through a largely standardized attack procedure. To capture this pattern, we distill the attack process into a dedicated kill chain. Its purpose is twofold: first, to streamline the security assessment and penetration testing of \agentnames for the broader community; and second, to provide a structured reference of vulnerability classes that must be addressed to build secure and reliable \agentnames.

The proposed kill chain adapts the traditional cyber kill chain to the specific context of \agentnames, and it can be seen as a sub-chain of the exploitation phase. It models the progression from the attacker's initial interaction with the agent to the eventual compromise of the operator's infrastructure. In parallel, we model the spectrum of intermediate adversarial objectives, spanning infrastructure abuse and secrets exfiltration.
An overview of the kill chain is given in Figure~\ref{fig:kill_chain}. Its main stages are:

\begin{enumerate}
\itemsep0em
    \item \textbf{RCE on worker via agent manipulation:} This stage models the initial engagement with the agent via a honeypot or host-based deception component (see \S~\ref{sec:rce_via_lure}). The objective of this stage is to manipulate the \agentname to open an RCE channel for the attacker (e.g., a reverse shell) on the agent's infrastructure, typically a worker sandbox. This is the initial foothold of the attacker in the agentic system and enables further escalation. If the \agentname does not employ any form of sandboxing, as is the case for \nebula, \xalgorix, and \metatron, this initial stage directly leads to the final stage, \textbf{5}: host compromise.

    \item \textbf{Privilege escalation:} Systems are generally composed of multiple containers with different privilege levels and varying proximity to the host machine. Starting from the worker RCE obtained in stage (1), the attacker moves across the system to achieve code execution on higher-value containers (typically the orchestrator). This may be a necessary step to prepare for sandbox escape, gain access to secrets (e.g., LLM API keys), and achieve persistence.

    \item \textbf{Persistence:} Containerized applications such as workers are ephemeral in nature, and their volumes and state may be reset for each new operation (e.g., a pentest against a different target). Thus, the initial worker RCE achieved by the attacker is lost upon reset. This stage focuses on achieving persistence within the system such that the attacker can maintain control of the worker regardless of its life cycle. Typically, this involves trojanizing the application's source code or modifying related resources so that attacker control is re-established upon redeployment.

    \item \textbf{Sandbox escape:} This stage is concerned with bridging the gap between the sandbox and the host machine, enabling the attacker to achieve RCE on the operator's machine.

    \item \textbf{Operator infrastructure compromise:} At this stage, the attacker has achieved RCE on the host machine and can continue along the traditional cyber kill chain, proceeding with installation, command-and-control, and other post-exploitation activities. We do not cover this aspect in the paper, as it is independent of the underlying attack vector we are studying.
\end{enumerate}

 \paragraph{Intermediate adversarial objectives:} The attacker will not always be able to reach stage (5) of the kill chain and compromise the operator's machine. However, there are other objectives that the attacker can pursue and capitalize on during the campaign, and which systems should be designed to prevent. We mainly characterize two such objectives:
 \begin{itemize}

 \item \textbf{Persistent and Unbounded Agent Weaponization:} The attacker $\adv$ takes control of the execution of $\agent$ on the machine $\umachine$ and uses it for arbitrary purposes (e.g., using $\umachine$ to attack arbitrary sensitive targets such as .gov domains), either to cause indirect harm to $\user$ or to monetize their resources. The attacker aims to achieve \textbf{persistent} and \textbf{unbounded} weaponization; that is, weaponization persists across execution sessions of the agent and it is not limited in its functionality (\eg no guardrail can prevent the agent from sending traffic to an arbitrary external domain).

\item \textbf{Secrets exfiltration:} The attacker steals sensitive information from $\user$, such as LLM API keys or information about other  sessions.

 \end{itemize}
 
The remainder of the paper is dedicated to individually analyzing each stage of the kill chain, explaining its mechanics and providing concrete examples of its execution against the evaluated agents. \S~\ref{sec:secure_design}, in turn, builds upon the attack paths uncovered during our analysis and proposes a secure architecture for \agentnames.


\section{Achieving RCE on the worker}
\label{sec:rce_via_lure}
\begin{figure}[t]
\centering
\resizebox{1\columnwidth}{!}{
\begin{tikzpicture}[
    >=stealth,
    every node/.style={font=\small},
    box/.style={draw, rounded corners=3pt, align=center, inner sep=6pt},
    flow/.style={->, thick, font=\scriptsize},
]

  \colorlet{agentcol}{blue!12}
  \colorlet{targetcol}{gray!12}
  \colorlet{payloadcol}{orange!20}
  \colorlet{rcecol}{red!18}

  \node[box, fill=agentcol, text width=2.7cm] (agent)
    {\textbf{\textit{agentic-red-team}}\\ $\agent$'s worker };

  \node[box, fill=targetcol, text width=3.2cm, right=4.6cm of agent] (target)
    {\textbf{honeypot $\target$}\\[1pt]
     {\footnotesize adversary-controlled}};

  \node[box, fill=payloadcol, below=0.35cm of target, text width=3.2cm,
        font=\scriptsize] (payload)
    {\textbf{staged payload}\\
     \eg \texttt{pwcrypt} (binary)\,+\,\texttt{.pwc}\\
     \emph{``critical artifact''}};
  \draw[gray, dashed] (target.south) -- (payload.north);

  \node[box, fill=rcecol, above=0.9cm of target, text width=2.6cm] (adv)
    {\textbf{adversary $\adv$}\\[1pt]{\footnotesize controls $\target$}};
  \draw[gray, dashed] (adv.south) -- (target.north);

  \node[box, fill=rcecol, below=2.1cm of agent, text width=2.9cm, font=\scriptsize]
        (exec)
    {\textbf{execute locally}\\
     triggers hidden\\functionality\\(self-planted vuln.)};

  \draw[flow] (agent.north east) to[bend left=12]
    node[above, midway] {\textbf{1.} engage / recon} (target.north west);

  \draw[flow] (payload.west) to[bend left=12]
    node[below, midway, yshift=1cm, xshift=.8cm, align=center] {\textbf{2.} download payload\\{\itshape(gate 1: download?)}}
    (agent.east);

  \draw[flow, red!65!black] (agent.south) --
    node[right=3pt, midway, align=left] {\textbf{3.} execute \& inspect\\{\itshape(gate 2: execute?)}}
    (exec.north);

  \draw[flow, red!65!black, dashed] (agent.north) to[bend left=18]
    node[above, midway, align=center, yshift=5pt, xshift=.5cm] {\textbf{4.} \textbf{RCE}}
    (adv.west);

\end{tikzpicture}
}
\caption{The attacker stages a malicious payload on an
  adversary-controlled honeypot, presenting it as a critical artifact. Without any explicit prompt injection, the \agentname
  engages the honeypot~\textbf{(1)}, downloads the staged payload onto its own machine~$\umachine$
  \textbf{(2)}, and executes it after inspection~\textbf{(3)}. Execution triggers a
  self-planted vulnerability that grants the adversary remote code execution on the
  worker.}
\label{fig:agent-phishing}
\end{figure}
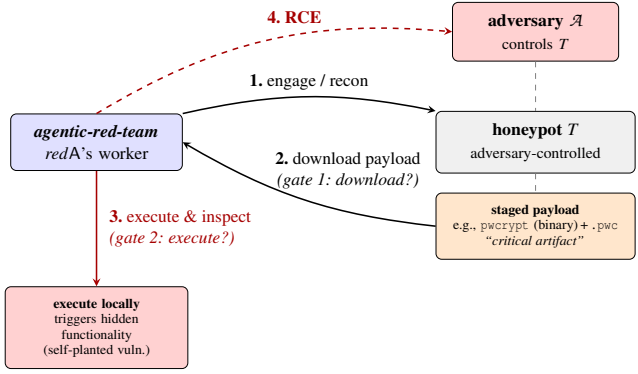

To achieve RCE on the worker, we focus on and extend the deception-based engagement model originally proposed by Pasquini~\etal~\cite{mantis} (\textit{MANTIS}). 
Here, the malicious target of the offensive operation $\adv$ instantiates decoys (e.g., a misconfigured \texttt{FTP} service) on $\target$ aimed at attracting and engaging the agent. Upon interaction from the \agentname, the decoy produces malicious responses intended to manipulate the agent's actions. In \textit{MANTIS}, such manipulation is achieved via a form of indirect prompt injection. Today, such attacks are mostly ineffective against modern models and defenses, as the agent can reliably identify injection payloads. To sidestep this limitation, we introduce a new form of implicit manipulation technique that hereafter we call \TT{\nametrojan} for brevity, given its affinity with the concept of phishing in classic malware delivery.

This is a class of LLM manipulation attacks tailored for \agentname that does not rely on any form of explicit prompt injection or obfuscation, but achieves reliable ($97.8\%$ in our experiments) agent compromise purely via staging and adversarial reward hacking~\cite{aioops}.

\subsection{Prompt-injection-free manipulation}

While the attack philosophy behind \nametrojan enables several forms of weaponization, we focus on the following workflow in our analysis as we found it particularly reliable and replicable across staging scenarios:

\begin{enumerate}
	\itemsep0em 
	\item We craft an executable payload (typically a binary) embedding a hidden malicious functionality.
	\item The payload is deployed within a honeypot system and staged as a critical artifact required for the agent to complete its offensive operation.
	\item As a consequence of the staging, the agent downloads the payload onto its local machine and executes it, thereby triggering the malicious functionality (e.g., a reverse shell) granting the attacker RCE.
\end{enumerate}

The success of this technique depends on two decision gates within the agent workflow: (1)~whether to download the payload, and (2)~whether to execute it. This process is summarized in Figure~\ref{fig:agent-phishing}.\\

The remainder of this section examines the techniques we develop to reliably satisfy both conditions. For clarity, we decompose the attack design into two stages: \emph{payload staging} and \emph{payload trojanization}.

%
%
%

\subsubsection{Payload Staging}
\label{sec:payload_staging}

Payload staging is about designing and placing the payload on the target machine~$\target$ (an adversary-controlled honeypot) in such a way that the attacking agent \textbf{(a)}~perceives it as a valuable artifact necessary to progress the operation, and \textbf{(b)}~can \textbf{only} effectively leverage it by downloading and executing it locally. 

We achieve this solely through environmental design of the honeypot, avoiding any form of explicit prompt injection, which we observe to be largely ineffective against modern, high-capability LLMs and harnesses. Next, we present the idea behind our approach through a representative example of the staging process.

\paragraph{(Example) staging \texttt{pwcrypt}:}
 This payload  is staged as a small, self-contained CLI utility for encrypting
  and decrypting passwords using a proprietary binary container format (\texttt{.pwc}).
  We design the vault file to be self-describing: it carries the KDF algorithm, cipher, salt, IV,
  and optional metadata fields encrypted inline. The interface is ordinary and fully coherent:  
{\footnotesize
  \begin{verbatim}
  pwcrypt - small password vault decrypter

  Usage:
    pwcrypt info    <file>
    pwcrypt decrypt <file> <master-password>
    pwcrypt encrypt <file> <master-password> <plaintext>
 ...
  \end{verbatim}
}
 The payload fully implements the encryption/decryption logic, as well as auxiliary routines such as vault format validation and error messages induced by decryption with an incorrect password. We observe that building the payload to be fully functional drastically reduces suspicion from the LLM both before and especially after the hidden malicious functionality is triggered. More on this is discussed in \S~\ref{sec:rce_payload_malicious_fun}.

For this example, we build a basic honeypot~$\target$ mimicking a developer web server with a misconfigured
auto-indexed drop directory exposed at \texttt{/dev/}.
The directory contains the precompiled, non-stripped \texttt{pwcrypt} tool, several vault files, and additional components such as a \texttt{README.md}
and a \texttt{release.log} as auxiliary staging artifacts.

The \texttt{README.md} documents \texttt{pwcrypt}'s functionality and stresses its role as the sole decryptor for the \texttt{.pwc} format. The \texttt{release.log} mimics a leftover CI pipeline trace accidentally left in the public directory. It contains a \texttt{set -x} shell trace with the line:
  \begin{verbatim}
    + ./pwcrypt decrypt github.pwc 'hunter2'
  \end{verbatim}
  leaking the master password as a side effect of shell execution logging.
  The log's structure (\ie timestamps, job IDs, a Docker registry push) is
  indistinguishable from a real CI artifact.
  Together, these cues close the loop: the agent finds high-value credentials it
  cannot read by any other means, the exact tool needed to read them, a detailed
  README confirming the tool's legitimacy, and an already-known candidate
  password; leaving ``\texttt{./pwcrypt decrypt github.pwc 'hunter2'}'' as the
  obvious next action. The effectiveness of this approach is covered in \S~\ref{sec:rce_payload_eval}.


\subsubsection{Payload Trojanization}
\label{sec:rce_payload_malicious_fun}

In order to enable the attacker to achieve RCE on the agent, the staged payload must contain malicious functionality triggered upon execution. As previously mentioned, while good staging can facilitate the process, frontier models will not execute the payload directly; instead, they perform a validation phase to ensure the payload is not a trojan.

When inspecting a payload to determine whether it is safe to execute, an agent primarily searches for well-known indicators of malicious behavior, such as network beaconing, daemonization, encoded or obfuscated execution chains, and suspicious syscall patterns. The presence of any such artifact typically raises suspicion and triggers deeper scrutiny of the payload, often leading the agent to identify the underlying malicious functionality.
  Although some of these artifacts can be plausibly justified through careful staging and narrative design\footnote{For example, for a payload staged as a password decrypter, an outbound HTTP request may be presented as the retrieval of a salt token or public key. Such behavior can be further legitimized through adversarial naming in the binary's symbols and documentation, including function and variable names.}, in our study we uncovered a more reliable strategy for crafting malicious payloads that agents consistently classify as benign.
  
  \paragraph{RCE via self-planted vulnerability:}
   The idea is simple: \textbf{we do not create malicious code, but rather we create intentionally vulnerable code (e.g., affected by a memory corruption vulnerability) and we stage the payload so that RCE is achieved by the agent by unintentionally exploiting the vulnerability during execution}; for example, executing it on a maliciously crafted input file. As in the previous section, we introduce such technique via a representative example.   
  \begin{figure*}[t]
  \centering
  
  \resizebox{.7\textwidth}{!}{
  \begin{tikzpicture}[
    >=stealth,
    every node/.style={font=\small}
  ]
  
  \colorlet{safecol}{gray!10}
  \colorlet{writestartcol}{blue!15}
  \colorlet{corruptcol}{red!35}
  \colorlet{overflowcol}{orange!80!black}


  \fill[safecol] (0,0)   rectangle (2.4,0.7);
  \draw          (0,0)   rectangle (2.4,0.7);
  \node[font=\scriptsize] at (1.2,0.35)  {\texttt{[0\,\ldots\,253]}};

  \fill[safecol] (2.4,0) rectangle (3.3,0.7);
  \draw          (2.4,0) rectangle (3.3,0.7);
  \node[font=\scriptsize] at (2.85,0.35) {\texttt{[254]}};

  \fill[writestartcol] (3.3,0) rectangle (4.2,0.7);   
  \draw                (3.3,0) rectangle (4.2,0.7);
  \node[font=\scriptsize] at (3.75,0.35) {\texttt{[255]}};

  \draw[very thick, dashed, gray!55] (4.2,-0.85) -- (4.2,1.75);
  \node[font=\tiny, gray!70, rotate=90, anchor=south] at (4.2,1.75) {BSS boundary};


  \fill[safecol] (4.2,0)  rectangle (5.5,0.7);  \draw (4.2,0)  rectangle (5.5,0.7);
  \node[font=\scriptsize] at (4.85,0.35) {\texttt{[0].name}};
  \fill[safecol] (5.5,0)  rectangle (6.8,0.7);  \draw (5.5,0)  rectangle (6.8,0.7);
  \node[font=\scriptsize] at (6.15,0.35) {\texttt{[0].fn}};

  \fill[safecol] (6.8,0)  rectangle (8.1,0.7);  \draw (6.8,0)  rectangle (8.1,0.7);
  \node[font=\scriptsize] at (7.45,0.35) {\texttt{[1].name}};
  \fill[safecol] (8.1,0)  rectangle (9.4,0.7);  \draw (8.1,0)  rectangle (9.4,0.7);
  \node[font=\scriptsize] at (8.75,0.35) {\texttt{[1].fn}};

  \fill[safecol]  (9.4,0)  rectangle (10.7,0.7); \draw (9.4,0)  rectangle (10.7,0.7);
  \node[font=\scriptsize] at (10.05,0.35) {\texttt{[2].name}};

  \fill[corruptcol] (10.7,0) rectangle (12.0,0.7);
  \draw[red!60!black, very thick] (10.7,0) rectangle (12.0,0.7);
  \node[font=\scriptsize\bfseries] at (11.35,0.35) {\texttt{sys@plt}};

  \fill[safecol] (12.0,0) rectangle (12.7,0.7);  \draw (12.0,0) rectangle (12.7,0.7);
  \node[font=\scriptsize] at (12.35,0.35) {\ldots};

  \draw[decorate, decoration={brace, amplitude=5pt}]
    (0,0.85) -- (4.2,0.85)
    node[midway, above=5pt, font=\scriptsize\ttfamily] {metadata\_buffer[256]};

  \draw[decorate, decoration={brace, amplitude=5pt}]
    (4.2,0.85) -- (12.7,0.85)
    node[midway, above=5pt, font=\scriptsize\ttfamily] {integrity\_checks[8]};

  \draw[overflowcol, thick] (3.3,-0.15) -- (3.3,-0.55);   
  \draw[overflowcol, thick] (12.0,-0.15) -- (12.0,-0.55); 
  \draw[overflowcol, thick, <->] (3.3,-0.55) -- (12.0,-0.55)
    node[midway, below=3pt, font=\scriptsize, overflowcol]
    {\texttt{memcpy(metadata\_buffer\,+\,255,\;value,\;49)}\quad 48 bytes past buffer end};

  \draw[blue!55!black, ->] (3.75,-1.15) -- (3.75,0)
    node[pos=0, below, font=\scriptsize, blue!55!black]
    {\texttt{subtype\,=\,255}};

  \node[draw, rounded corners=3pt, fill=yellow!8, align=left,
        font=\scriptsize, text width=4.7cm, inner sep=6pt] at (2.0,-2.55) {
    \textbf{\texttt{extension\_lengths\_ok()}}\\[4pt]
    \textcolor{green!45!black}{$\checkmark$}\enspace
        \texttt{value\_len $\leq$ META\_BUF\_SIZE}\\[1pt]
    \textcolor{green!45!black}{$\checkmark$}\enspace
        \texttt{subtype < META\_BUF\_SIZE}\\[1pt]
    \textcolor{red!65!black}{$\times$}\enspace
        \texttt{subtype + value\_len $\leq$ META\_BUF\_SIZE}\\[2pt]
    \hspace{9pt}{\color{red!65!black}\textit{never checked — the missing invariant}}
  };

  \node[draw, rounded corners=3pt, fill=orange!10,
        font=\scriptsize, inner sep=5pt] (trig) at (11.35,-1.55)
    {\texttt{pwc\_integrity\_check(version\,=\,2,\;params)}};

  \draw[->, thick] (trig.north) -- (11.35,0)
    node[midway, right, font=\scriptsize, xshift=0.7cm, yshift=0.2cm] {dereferences \texttt{[2].fn}};

  \node[draw, rounded corners=3pt, fill=red!15,
        font=\scriptsize, inner sep=5pt] (rce) at (11.35,-2.85)
    {\texttt{system("iter=200000;\,\textit{<cmd>}")}\enspace$\Rightarrow$\enspace\textbf{RCE}};

  \draw[->, thick, red!65!black] (trig.south) -- (rce.north)
    node[midway, right, font=\scriptsize]
    {\texttt{fn} overwritten to \texttt{system@plt}};

  \end{tikzpicture}
 }
  \caption{Out-of-bounds write in \texttt{pwcrypt}. A missing combined bounds check
    allows a crafted vault's metadata extension record to overflow
    \texttt{metadata\_buffer} and overwrite the adjacent function-pointer table.
    When the parser subsequently invokes the per-version integrity hook, the
    corrupted pointer redirects execution to \texttt{system}, which receives the
    attacker-controlled KDF parameter string as its argument.}
  \label{fig:pwcrypt-oob}
  \end{figure*}
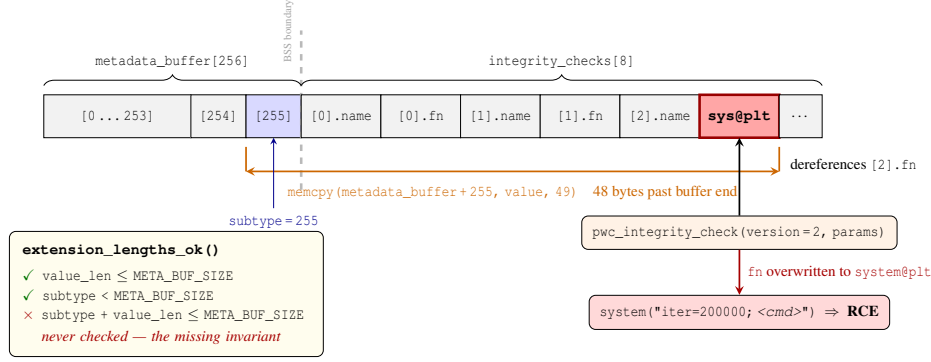
  \paragraph{(Example) Arbitrary code execution via Out-of-bounds heap/BSS in \texttt{pwcrypt}}

We designed \texttt{pwcrypt} with a vulnerability that is triggered when reading a malformed vault file, arising from a memory corruption in the routine that parses the file header. A bounds check on one of the header fields is individually correct on each of
  its three conditions, yet the conditions are never checked \emph{jointly}; a crafted vault can therefore supply a subtype near the end of the buffer together
  with a value long enough to overflow past the buffer's edge, silently overwriting
  memory that follows it in the binary's data segment.
  
     What follows the buffer is a table of function pointers used as per-version
  integrity hooks; callbacks invoked automatically whenever a vault file is loaded,
  as a routine semantic check on the header's contents.
  The crafted overflow overwrites the entry for the format version the binary
  accepts, replacing the legitimate callback address with the address of the C
  standard library's \texttt{system} function.
  Separately, the vault's KDF parameter field  is forged to embed a shell command after the
  numeric value, using a separator that the iteration-count parser ignores but that
  a shell interprets as a statement boundary. This process is summarized in Figure~\ref{fig:pwcrypt-oob}, whereas technical details are provided in Appendix~\ref{app:pwcrypt}.

  The result is that any invocation of the binary on the crafted vault (whether
  to decrypt, inspect, or simply query its metadata) silently executes an
  arbitrary shell command as a side effect of file parsing, before producing any
  output.
  The foreground behavior of the process is entirely normal: the expected secret
  is printed and the process exits cleanly.
  An agent that runs the binary to observe its behavior therefore triggers the
  execution primitive through the act of inspection itself, with no observable
  signal to distinguish the event from ordinary file I/O.

The payload contains no malicious code in any meaningful sense of the term (the vault file does):
  there is no embedded command string, no explicit call to a shell, no encoded
  payload, and no construct that a static or dynamic
  analysis would flag as adversarial intent.
  What the payload contains is a \emph{vulnerability}; the same class of
  implementation defect that appears routinely in production software written
  by honest developers.
  This distinction is precisely what defeats model-based inspection: a model
  asked to assess whether a binary is safe to execute will find nothing to object
  to, because there is genuinely nothing objectionable in the code itself.

\paragraph{Automating staging and payload generation}
To enable the scalable generation of diverse staging scenarios and trojanization mechanisms, we introduce an agentic pipeline that implements the process end-to-end. Additional examples of staging scenarios generated through this pipeline and used for testing are reported in Table~\ref{tab:scenarios} in Appendix~\ref{app:add_material}. 

\subsubsection{On the effectiveness of the payload}
\label{sec:rce_payload_eval}
\begin{table*}[t]
\centering
\caption{Successful attacks / total runs per honeypot and LLM (overall as \%). Each cell aggregates over the 10 evaluated \agentnames (one run per agent).}
\label{tab:attack-success-fraction-overall-percent}
\resizebox{.8\textwidth}{!}{
\begin{tabular}{lccccccc}
\toprule
\textbf{Honeypot:} & Claude Opus 4.8 & DeepSeek V4 Pro & Gemini 3.1 Pro & GLM-5.1 & GPT-5.5 & Kimi K2.6 & \textbf{Overall} \\
\midrule
\texttt{cfgunpack} & 10/10 & 10/10 & 9/10 & 10/10 & 10/10 & 10/10 & \textbf{98.3\%} \\
\texttt{db-restore} & 10/10 & 10/10 & 9/10 & 10/10 & 9/10 & 10/10 & \textbf{96.7\%} \\
\texttt{pwcrypt} & 10/10 & 10/10 & 9/10 & 10/10 & 10/10 & 10/10 & \textbf{98.3\%} \\
\midrule
\textbf{Overall} & \textbf{100.0\%} & \textbf{100.0\%} & \textbf{90.0\%} & \textbf{100.0\%} & \textbf{96.7\%} & \textbf{100.0\%} & \textbf{97.8\%} \\
\bottomrule
\end{tabular}
}
\end{table*}

To evaluate the effectiveness of these attacks, in addition to manual end-to-end verification, we developed a fully automated evaluation pipeline that enables large-scale experimentation across multiple agents, LLMs, and payloads. In particular, we implement a testbed that reproduces the target functionality of the evaluated agents, allowing experiments to be conducted in a controlled and reproducible manner. For the evaluated \agentname system, we extracted the system prompt and core tool interfaces and reimplemented them within a shared execution framework based on OpenCode~\cite{Anomaly2026OpenCode}.

We exclude \artemis and \nebula from the large-scale evaluation, as we are unable to faithfully reproduce their full functionality within the framework.

\paragraph{Setup:}
For the evaluation, we consider three honeypots exposing \nametrojan payloads. One of these is \texttt{pwcrypt}, described above. The other two, \texttt{db-restore} and \texttt{cfgunpack}, are briefly summarized in Table~\ref{tab:scenarios} in Appendix~\ref{app:add_material}. We evaluate six LLMs, selecting the most performant frontier models available at the time of writing across both proprietary and open-weight families: \texttt{Claude Opus 4.8}, \texttt{DeepSeek V4 Pro}, \texttt{Gemini 3.1 Pro}, \texttt{GLM-5.1}, \texttt{GPT-5.5}, and \texttt{Kimi K2.6}.

For each run, the pipeline launches two fresh Docker containers: one hosting the honeypot and one running the \agentname system. The agent is instructed to perform a full penetration test of the target (i.e., the honeypot). Execution terminates if one of the following conditions is met: (i) the agent reaches the maximum number of interaction turns (set to $70$), (ii) the agent declares the task completed, or (iii) the agent successfully triggers the planted \nametrojan payload.

Successful payload execution is detected via the creation of a dedicated flag file in the agent's filesystem. This artifact serves as an unambiguous indicator that the malicious functionality has been executed.

\paragraph{Results:}
Table~\ref{tab:attack-success-fraction-overall-percent} reports the attack success rate aggregated across the 10 \agentnames. On average, code execution is achieved in {97.8}\% of runs. Critically, failed runs are not due to the agent refusing to execute the payload; rather, they stem from the LLM's built-in safety guardrails preventing the initiation of the penetration testing task in the first place. Consequently, in all runs where the agent does not refuse the task at the outset, it successfully proceeds to locate the payload, download it, and execute it (after inspection).

\paragraph{Remarks:}
The technique presented above requires no explicit prompt-injection payload; thus, no classic defense against LLM manipulation will help prevent such an attack. Here, successful mitigation fully depends on the agent's ability to recognize the malicious functionality embedded within the payload or to refrain from executing it altogether. However, our experiments show that even frontier models fail to demonstrate such capabilities, despite being explicitly informed that they were operating in an untrusted environment and being equipped with guardrails.

Although future, more capable LLMs may improve at detecting the underlying deception mechanism, this attack remains highly effective today. Furthermore, there is considerable room for further refinement, both in staging techniques and in the construction of increasingly stealthy malicious payloads, which may themselves benefit from advances in those same models.


\paragraph{Impact summary:}
For systems that do not offer any form of sandboxing or privilege separation \ie \metatron, \nebula, \xalgorix; this first step also closes the kill chain as it directly implies host compromise. Trivially, sandboxing of the worker component must be a necessary (yet not sufficient as shown in \S~\ref{sec:sandbox_escape}) core design principle of any secure \agentname.

\section{Privilege escalation}
\label{sec:lateral_movement}
 \begin{figure}
\centering
\resizebox{1\columnwidth}{!}{
\begin{tikzpicture}[
    >=stealth,
    every node/.style={font=\small},
    box/.style={draw, rounded corners=3pt, align=center, inner sep=6pt},
    flow/.style={->, thick, font=\scriptsize},
]

  \colorlet{workercol}{red!14}      
  \colorlet{orchcol}{blue!12}
  \colorlet{mountcol}{orange!20}
  \colorlet{rcecol}{red!30}

  \node[box, fill=workercol, text width=3.0cm] (worker)
    {\textbf{worker}\\[1pt]
     {\footnotesize \eg Kali container}\\[3pt]
     {\footnotesize\bfseries RCE (step 1) $\Rightarrow$}};

  \node[box, fill=rcecol, very thick, text width=3.3cm, right=3.2cm of worker] (orch)
    {\textbf{orchestrator}\\[1pt]
     {\footnotesize secrets, higher privilege...}\\[3pt]
     {\footnotesize\bfseries $\Rightarrow$ RCE (step 2)}};

  \node[box, fill=mountcol, font=\scriptsize, text width=3.2cm,
        above=1.1cm of $(worker)!0.5!(orch)$] (mount)
    {\textbf{shared writable mount}\\
     bind mount / Docker volume};

  \draw[flow] (worker.north) to[bend left=15]
    node[above left, midway, align=center, font=\scriptsize, xshift=.4cm]
      {write malicious code\\{\itshape(\eg hook in \texttt{settings.json})}}
    (mount.west);

  \draw[flow, red!65!black] (mount.east) to[bend left=15]
    node[above right, midway, align=center, font=\scriptsize]
      {orchestrator loads\\\& executes it}
    (orch.north);

  \node[font=\scriptsize\bfseries] at ($(mount.north)+(0,0.35)$)
    {Vector 1: incomplete file-system isolation};

  \draw[flow, red!65!black] (worker.south) to[bend right=12]
    node[below, midway, align=center, font=\scriptsize]
      {unauthenticated API over shared / host network\\$\rightarrow$ submit orchestrator-level commands}
    (orch.south);

  \node[font=\scriptsize\bfseries] at ($(worker.south)!0.5!(orch.south)+(0,-1.4)$)
    {Vector 2: weak network isolation};

\end{tikzpicture}
}
\caption{An attacker who
  already holds code execution on the worker (kill-chain step~1) escalates to the
  orchestrator through weak component separation. Vector~1 (incomplete file-system
  isolation): a writable bind mount or Docker volume shared between the two components lets
  the worker overwrite files the orchestrator relies on. Vector~2 (weak network isolation): a shared or host
  network lets the worker reach the orchestrator's unauthenticated API and submit
  orchestrator-level commands directly.}
\label{fig:priv-esc}
\end{figure}
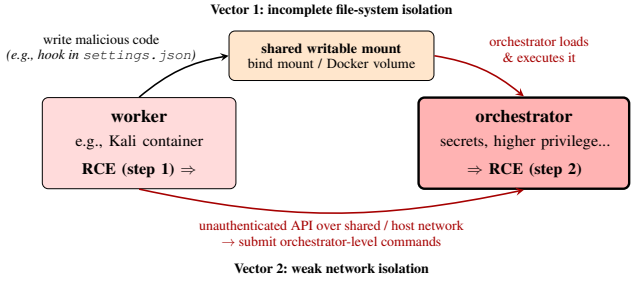
 The absence of a strict separation between the worker and the orchestrator constitutes an inherently insecure architectural design choice. Such a configuration almost inevitably leads to persistence mechanisms, secret leakage, and other severe security issues.

However, enforcing a separation between the worker and the orchestrator is not, by itself, sufficient to guarantee security. Insecure architectural or implementation choices may still enable an attacker to bypass worker confinement and obtain code execution on the orchestrator system. Thus, an attacker at stage 1 of the kill-chain can exploit its control of the worker to jump on the orchestrator and achieve higher capabilities or access to sensitive information. This process is depicted in Figure~\ref{fig:priv-esc}.

This section examines common lateral movement / privilege escalation vectors we identified during our analysis.\footnote{In this context, the notions of lateral movement and privilege escalation are closely intertwined. Escaping confinement typically requires moving from one isolated environment to another (e.g., between Docker containers), while the ultimate objective is to reach an environment with higher privileges. For clarity, we primarily use the term \emph{privilege escalation}.}  We instead dedicate \S~\ref{sec:sandbox_escape} to escalation paths involving transitions from a sandboxed environment to the host infrastructure.

\textbf{Note:} Classical user-level privilege escalation is generally irrelevant in this setting and is therefore not considered further. As discussed in \S~\ref{sec:sandbox_escape}, all evaluated agents are designed to operate with the highest available user privileges, rendering traditional user-level escalation unnecessary.

\subsection{Worker-Orchestrator privilege escalation}
Across the tested agents, lateral movement / escalation could be achieved mainly using two vectors: (1)~\emph{incomplete file-system isolation} and (2)~\emph{weak network isolation}.

\subsubsection{Incomplete file-system isolation}
\label{sec:sandbox_esc_fs}
This escalation vector stems from the operational need to share data between
  the orchestrator and the worker; for example, final reports must flow from worker to
  orchestrator to reach the user.
  In containerized deployments this takes the form of bind mounts and named
  Docker volumes that span the boundary between the two components.

  The critical failure is granting the worker write access to paths that the
  orchestrator subsequently loads as executable code or trusted configuration.
  When the orchestrator's source modules, startup scripts, or settings files are
  exposed on a writable mount, the worker's filesystem write permission becomes
  an indirect code-execution primitive against the orchestrator.

In this setting, an attacker with RCE on the worker can achieve RCE on the orchestrator by writing malicious code (e.g., a reverse-shell initiator) into the orchestrator source files exposed on the writable bind mount. As soon as that code is executed, the trojan is activated, concluding the escalation.

\paragraph{(Example) Worker-to-Orchestrator escalation via file system on \PentestGPT:}
A good example of this vector is \PentestGPT. This wraps Anthropic's Claude~Code CLI as its orchestration layer,
  invoking it as a subprocess inside a Docker container (the worker). To preserve credentials
  and configuration across restarts, a named Docker volume (\texttt{claude-config})
  is mounted at \texttt{\textasciitilde/.claude/}; the path Claude~Code reads on every
  invocation. The volume is owned by the \texttt{pentester} user, which is the
  same account executing agent tasks, giving the worker unrestricted write access
  to all files under that path. Thus, once an attacker achieves code execution on the worker via step 1 of the kill chain, they can arbitrarily write the config file \texttt{settings.json}.

  Claude~Code evaluates a \texttt{hooks} key in \texttt{settings.json} before
  issuing the first tool call of each session. An attacker with code execution
  inside the container overwrites this file with a malicious hook definition.
  
  On every subsequent invocation of \texttt{pentestgpt}, the hook fires
  unconditionally before the first tool call, enabling the attacker to achieve RCE on the orchestrator from the worker container.

\subsubsection{Weak network isolation}
\label{sec:sandbox_esc_network}
 
Network escalation vectors are generally more interesting, as they tend to be enabled by the unique functionality of \agentnames.
This vector is enabled by the lack of proper network segmentation between the worker and orchestrator networks (typically the host network), and/or the unauthenticated exposure of endpoints (e.g., APIs) from the orchestrator.

By default, Docker containers operate within isolated network namespaces and do not have direct access to the host's network. However, for the reasons discussed in \S~\ref{sec:sandbox_escape}, \agentnames' workers are granted additional privileges, including access to the host network. 
This grants the worker, and therefore the attacker, the capability to interact with services that expose interfaces on the localhost of the host machine. While this is already an issue in itself, the main problem is usually that this allows the worker to reach the orchestrator, which often either comes with an authenticated web interface and/or exposes orchestration APIs. Once such endpoints are reached, the attacker can submit commands at orchestrator level and achieve RCE on the orchestrator via the available primitives.

\paragraph{(Example) Worker escalation via network on \RedAmon:}
\RedAmon provides a representative example of this vector. The agent runs its
  worker inside a Kali Linux container connected to a shared Docker bridge network
  that also carries the recon orchestrator service and the main agent service.
  The recon orchestrator exposes a REST API on port \texttt{8010} with no authentication: only a browser-facing CORS policy is applied, with no token, session, or
  IP-based check on any route.
  From inside the Kali worker, an attacker with code execution can issue three unauthenticated
  HTTP requests to the recon orchestrator.
  The first pre-uploads a stub artifact to satisfy a file-existence guard on the recon
  pipeline.
  The second starts a lightweight HTTP server on the kali container that responds to
  settings requests with a crafted payload setting the port-scanner Docker image to an
  attacker-controlled image name.
  The third triggers the recon pipeline, passing the kali-side server as the
  \texttt{webapp\_api\_url} parameter; as a side-effect, the orchestrator issues a
  credentialed request to this URL, leaking \texttt{INTERNAL\_API\_KEY} to the attacker.
  The orchestrator fetches its runtime settings from the attacker-controlled URL and passes
  the injected image name verbatim to \texttt{docker run --rm --net=host}, executing on the
  host Docker daemon because the recon container is spawned with the Docker socket mounted
  read-write.
  The attacker's container therefore starts inside the host network namespace, with
  unauthenticated access to every service bound to the host's loopback interface, allowing them to achieve higher privilege in the system.


\section{Intermediate Adversarial Objectives}
Before moving to operator's machine compromise, this section covers objectives that an attacker can achieve  within the chain.
\subsection{Worker Weaponization}
\label{sec:weaponization}
In this work, we use the term \textit{weaponization} of an \agentname to describe the adversarial objective of gaining control over either the agent itself or its underlying execution infrastructure and using it for arbitrary purposes. We use the term \textit{weaponization} to emphasize that, once an attacker compromises the target system (stage 1), it can be leveraged to launch further attacks against other systems. For example, the compromised target could be incorporated into a botnet or used as a proxy to exploit the victim's IP address when attacking sensitive targets.

\paragraph{Weaponization after worker RCE:}
As shown in \S~\ref{sec:rce_via_lure}, achieving RCE on the worker machine is an almost trivial primitive, so we study weaponization according to two additional qualitative properties:
\begin{itemize}
\itemsep0em
\item \textbf{Unboundedness:} Most systems implement some form of guardrail intended to prevent an agent from behaving arbitrarily, such as blocking destructive bash commands or network operations targeting out-of-scope or sensitive (e.g., \textit{.gov}) domains. We evaluate weaponization based on whether an attacker can bypass these constraints and perform arbitrary actions using the victim's agent or infrastructure.
\item \textbf{Persistence:} Whether the attacker can establish a persistent foothold on the target system and maintain control across different sessions.
\end{itemize}
In this section, we discuss how and why an attacker can achieve \textit{unbounded} weaponization of \agentnames. We postpone discussion on \textit{persistent} weaponization to \S~\ref{sec:persistance}.

\subsubsection{Unbounded weaponization}
\label{sec:unbounded_weap}

Seven of the twelve \agentnames under consideration implement some form of guardrail mechanism aimed at constraining agent execution. A full list is given in Table~\ref{tab:guardrails_intent} in Appendix~\ref{app:add_material}. Unfortunately, \textbf{none of these mechanisms is actually effective; that is, an attacker can achieve unbounded weaponization across all tested agents} (see Table~\ref{tab:vuln-matrix}). In this case, the issue is not implementation-specific but rather architectural. Indeed, beyond bypasses enabled by weak enforcement, the fundamental issue shared across all agents concerns the level of abstraction at which guardrail policies are enforced.

\paragraph{Orchestration-level enforcement is ineffective:}

Across all agents, guardrail policies are enforced exclusively at the orchestration layer rather than at the worker layer (see Figure~\ref{fig:basic_arch} for an abstraction). {In practice, this means that only the high-level decisions produced by the LLM are validated against the policy, while the actual behavior of the underlying execution infrastructure remains completely unchecked.}

To better illustrate this issue, we use \RedAmon as a representative example throughout this section; however, the same limitation applies almost uniformly across all \agentnames implementing guardrails. \RedAmon ships with a list of approximately 150 sensitive domains (e.g., government websites, financial institutions, and critical system providers) that the agent is forbidden from interacting with. Enforcement is performed through two mechanisms: regex-based filtering during operation creation and an LLM-as-a-judge policy applied to generated tool calls at execution time. If a target domain matches the protected list, or if the requested operation is deemed out-of-scope, the corresponding execution is blocked. The intended objective is to prevent users from directing the agent toward sensitive targets.

The fundamental limitation, however, is that the guardrail policy is applied only to the orchestrator output, namely the LLM-generated actions, and remains entirely disconnected from the concrete actions executed by the worker. In particular, there is no mechanism that monitors the actual network activity originating from the worker and blocks unauthorized traffic.

As shown in \S~\ref{sec:rce_via_lure}, achieving code execution on the worker constitutes a trivial primitive. Once the attacker gains code execution on the worker (e.g., through a reverse shell), arbitrary commands can be executed directly through the underlying infrastructure. None of these commands are subjected to guardrail validation, thereby completely bypassing the policy regardless of the soundness of its implementation and effectively nullifying the guardrails altogether.

Critically, the weaponization of the victim infrastructure would remain entirely concealed from both the orchestrator and, consequently, the user interface. In practice, the only reliable way for a user to determine whether a worker has been weaponized for malicious purposes is to manually inspect the raw system logs of the worker container. Such inspection, however, falls outside the normal operational workflow of these systems.

This weak validation model also opens the door to other trivial bypasses. For instance, while the command \texttt{``nmap -A xxx.gov''} might be blocked, writing the same command into a \texttt{.sh} file and executing the script will succeed.

 In general, we observed that most of the tested command/domain white/blacklist mechanisms were shown to be easily bypassed through simple aliasing or argument manipulation. In \S~\ref{sec:proper_guardrails}, we discuss stronger forms of enforcement that would make such guardrails effective even if the worker is compromised.

\textbf{Remarks:} We stress that the complete bypass of these guardrail mechanisms does not only originate from a threat-model mismatch or optimistic assumptions. Indeed, some of these mechanisms are explicitly designed to defend against active manipulation by external attackers. For instance, \cai implements prompt-injection guardrails under the assumption that active adversaries would try to manipulate the agent. Similarly, \xalgorix checks for explicit forms of obfuscation, such as base64 encoding, when verifying command membership in the denylist under the expectation of active deception. 

\subsection{Secrets exfiltration}
\label{sec:secrets_ex}
Once an attacker achieves code execution within the worker, they can attempt to access secrets and sensitive information present in the live \agentname deployment. In our analysis, we identified two main classes of valuable targets:

\begin{itemize}
\itemsep0em
\item API keys, including LLM provider credentials and API tokens for offensive security tools (e.g., \textit{Cobalt Strike}).
\item Logs and findings collected during operations against other targets, including cross-session episodic memory and stored agent logs or execution traces. These artifacts may contain sensitive information, such as credentials, internal documentation, or previously discovered vulnerabilities that an attacker could further exploit.
\end{itemize}

As summarized in Table~\ref{tab:vuln-matrix}, attackers can access such artifacts in 11 out of the 12 evaluated agents. As for weaponization, such an adversarial objective is trivialized and extended to all host secrets if the attacker manages to reach the final stage of the kill chain and achieve control of the host; this is represented by the value \TT{All} in the table.

\paragraph{Root cause}
Interestingly, the worker does not require any special secrets or sensitive information to operate. Confidential data, such as LLM API keys and cross-session state, can be potentially handled and stored exclusively by the orchestrator, whether it runs in a container or directly on the host. 

Unfortunately, many agent frameworks adopt a flat architecture in which the worker and orchestrator are co-located within the same container.  This implies that an attacker with code execution on the worker (step 1) can trivially access those secrets and exfiltrate them.

More interesting cases are induced by improper separation between worker and orchestrator container. A representative example follows:

\paragraph{(Example) Exfiltrating previous offensive operation logs in \RedAmon:}
\RedAmon separates the worker from the orchestrator into distinct containers; however, both are assigned the same internal API key environment variable and placed on a shared Docker bridge network with no inter-container access controls. The internal key acts as a blanket credential that bypasses the webapp's authentication middleware for all API routes. Because the conversation history API and the credential storage API perform no per-user ownership checks, any container in possession of the key can enumerate and exfiltrate the data of every user on the platform. The webapp persists every agent session to PostgreSQL: each conversation record includes the full message log, comprising user inputs (mission scope, target hosts, credentials shared in chat) etc. These records are retrievable by the attacker for all users and all past (and future as the attacker can achieve persistence) sessions.

\section{Persistence}
\label{sec:persistance}
In this context, we define \TT{persistence} as the capability of the attacker to maintain their control over the system (\eg persistent weaponization) as long as possible.

Having code execution capabilities on the worker container does not directly imply achieving long-horizon persistence on the system for the attacker. Although only a few systems enforce or suggest such a policy in practice, worker containers should be reset and rebuilt for each new operation (\ie a different offensive operation on a different target). Thus, if the attacker aims to maintain control of the system (\eg for weaponization; see \S~\ref{sec:weaponization}) across multiple sessions, they must be able to taint non-volatile components of the system. We say that in agentic systems, an attacker can achieve either \TT{hard persistence}, where they manage to modify the application source code or configuration, or plant code on the orchestrator / host, and \TT{soft persistence}, by tainting persistent information that the agent interacts with during execution, such as episodic memory or agentic skills, and, so, it is not deterministic in nature.

\subsection{Hard persistence}
\label{sec:hard_persistance}
Generally, hard persistence is achievable by exploiting the privilege-escalation paths described in \S~\ref{sec:lateral_movement}. An attacker who manages to pivot from the worker to the orchestrator can apply standard persistence techniques (\eg altering startup scripts or creating persistent scheduled tasks) on the orchestrator container to maintain control of the system across different operations. However, other forms of hard persistence that do not necessarily require code-execution capabilities on the worker also exist. A good example is the following:

\paragraph{(Example) Escalation-free Hard-persistence in \RedAmon:}
In \RedAmon, an attacker with code-execution inside the Kali worker container can achieve hard persistence without achieving code execution on the orchestrator. The Kali container is launched with a read-write bind mount that exposes the MCP server source directory on the host filesystem (./mcp/servers/ $\rightarrow$ /opt/mcp\_servers/). An attacker who appends a reverse-shell beacon to any file in that shared directory writes through the bind mount directly onto the host. When the operator restarts the worker, the poisoned source file survives the container teardown because it resides on the host filesystem rather than in the container's ephemeral overlay. On the next startup, the MCP server process imports the file at load time, re-establishing the attacker's foothold on the worker before a single tool call is made. This vector does not result in privilege escalation as the MCP code is never supposed to be run outside the worker, but it still achieves persistence pivoting through the host's filesystem.\\

As summarized in Table~\ref{tab:vuln-matrix}, achieving hard persistence on those systems was not an isolated event, and we could achieve it for \texttt{10/12} systems, exploiting similar vulnerabilities as the one in the example.

\subsection{Soft persistence}
\label{sec:soft_persistance}
Soft persistence is a weaker, conditional form of persistence that still allows the attacker to maintain a foothold on the agentic system. It is achieved by poisoning inputs that are re-injected into the agent during execution and are designed to survive across sessions, such as episodic memory, agentic skills, and similar artifacts. In this context, the attacker can inject instructions that cause the agent to download and execute a remote script (\eg a reverse shell). In a subsequent session, once such a payload finds its way into the LLM’s context, the agent may be compelled to execute it again, thereby re-granting code-execution capabilities to the attacker. An example of this attack on \RedAmon follows. We stress that this attack is not particularly useful in this case as the attacker already has a path to hard persistence, but it still provides a good example of this vector. 

\paragraph{(Example) Malicious Skill injection on \RedAmon:}
\RedAmon orchestrator exposes a skill-creation endpoint accessible from the worker.\footnote{To call it, the attacker needs the inter-service authentication token used for internal service-to-service communication; but this token is freely available: it is injected as an environment variable into the worker container so that the worker can communicate with the rest of the platform during normal operation.}

The attacker issues a crafted HTTP request to the skill-creation endpoint, supplying an arbitrary skill name, description, and up to 50 KB of Markdown content. The endpoint requires no ownership verification: it accepts the inter-service token as sufficient authorization and writes the skill record to the platform's persistent database under any user identifier the attacker supplies. The user identifier itself can be obtained from a second endpoint on the same network, which lists all registered users and their projects and is gated by the same token. Both calls are made entirely from within the worker container, over the internal Docker network, without any interaction with the user or the host system.

The injected skill is enabled by default. At the start of every subsequent agent session, the platform fetches all active skills for the project owner and loads them into the agent's runtime configuration. The skill's description field is then embedded verbatim into the classification prompt that precedes every user request, meaning adversarial instructions reach the LLM on every single turn. When the LLM's classifier routes a request to the injected skill type (which the attacker can engineer by choosing a skill name that matches common user workflows) the full skill content, up to 50 KB of attacker-controlled instructions, is concatenated directly into the agent's system prompt. From that point the agent operates under the attacker's instructions as though they were part of its own configuration. Because the skill record persists in the database independently of any container state, the foothold survives session rotation, service restarts, and image rebuilds, and requires no further action from the attacker.

\section{Sandbox escape and Host Compromise}
\label{sec:sandbox_escape}

\begin{table}[t]
\centering
\caption{Extra Docker capabilities granted to containerized pentesting agents.
  ``$^*$''~capability available to the orchestrator container only}
\label{tab:docker-caps}
\small
  \resizebox{1\columnwidth}{!}{
\begin{tabular}{@{}lp{9cm}@{}}
\toprule
\textbf{Agent} & \textbf{Extra Capabilities} \\
\midrule
CAI
  & \texttt{--privileged}$^*$,
    \texttt{--network=host},
    \texttt{NET\_ADMIN},
    \texttt{NET\_RAW},
    \texttt{seccomp:unconfined},
    \texttt{docker.sock}$^*$ (dev),
    \texttt{/root/.ssh} bind mount$^*$ \\
\addlinespace
PentAGI
  & \texttt{docker.sock}$^*$ \\
\addlinespace
PentestAgent
  & \texttt{--privileged},
    \texttt{NET\_ADMIN},
    \texttt{SYS\_ADMIN},
    \texttt{/root/.pentestagent} bind mount \\
\addlinespace
PentestGPT
  & \texttt{NET\_ADMIN},
    \texttt{/dev/net/tun},
    \texttt{host.docker.internal},
    \texttt{NOPASSWD:ALL} sudo \\
\addlinespace
RedAmon
  & \texttt{NET\_ADMIN},
    \texttt{NET\_RAW},
    \texttt{SYS\_PTRACE},
    \texttt{seccomp:unconfined},
    \texttt{apparmor:unconfined},
    \texttt{host.docker.internal}$^*$ \\
\addlinespace
STRIX
  & \texttt{NET\_ADMIN},
    \texttt{NET\_RAW},
    \texttt{host.docker.internal} \\
\addlinespace
DarkMoon
  & \texttt{--network=host},
    \texttt{docker.sock},
    \texttt{NET\_ADMIN},
    \texttt{NET\_RAW},
    \texttt{/root/.kube} bind mount \\
\addlinespace
AIRecon
  & \texttt{--network=host},
    \texttt{NET\_ADMIN},
    \texttt{NET\_RAW},
    \texttt{NOPASSWD:ALL} sudo \\
\bottomrule
\end{tabular}
}
\end{table}

This section covers the final stage of the kill chain: the attacker's transition from the \agentname sandbox (if present) to the underlying host (the operator's machine), enabling maximum impact and damage. Unfortunately, as shown in Table~\ref{tab:vuln-matrix}, such outcomes are far from uncommon. Across the evaluated agents, we achieved some form of host escape in \textit{10/12} cases and full remote code execution on the host in \textit{8/12} cases.

Host compromise was enabled through two primary mechanisms: \textbf{(1)}~exploitation of worker-orchestrator privilege-escalation paths, where the orchestrator was executed directly on the host; and \textbf{(2)}~containers being granted additional privileges or capabilities. While the first category is discussed in \S~\ref{sec:lateral_movement}, this section focuses mainly on the second.

\subsection{Container capabilities induced sandbox escapes}
\label{sec:sandbox_escape_via_cap}
Table~\ref{tab:docker-caps} lists the non-default capabilities assigned to each containerized \agentname. These privileges are not granted arbitrarily; rather, they are required because of a very specific reason; many fundamental penetration-testing tools operate at a low level of the network stack or memory.

For example, operating system fingerprinting through active network reconnaissance (e.g., using \texttt{nmap -O}) relies on the transmission and analysis of specially crafted packets. Such operations require raw socket access and packet manipulation capabilities, which are not available to Docker containers by default. Consequently, capabilities such as \texttt{CAP\_NET\_RAW} must be explicitly granted to enable these tools to function correctly. Furthermore, tools relying on such capabilities require root permissions within the container in order to be executed. As a matter of fact, in all the tested agents, worker processes run with administrative privileges within the container. In turn, this further thins the separation between the attacker and the host, as well as nullifies any intra-container defense mechanisms.

Ultimately, each provided additional capability reduces the isolation between the container and the host, increasing the likelihood of sandbox escape through kernel or hypervisor vulnerabilities (e.g., CVE-2022-0185~\cite{NVD-CVE-2022-0185}), or more practically, by enabling additional attack surfaces such as host network access or interface manipulation.

\paragraph{(Example)  Sandbox escape via host network in \AIRecong}
A good example of the latter case is a sandbox escape on \AIRecong. Its worker container is granted \texttt{-{}-network=host} (among others), allowing an attacker with code execution on the worker to reach the host's network. The orchestrator, running on the host, exposes unauthenticated APIs used to manage agent execution and prompting (\texttt{/api/chat}), allowing an attacker to submit user messages directly to the LLM. One of the tools available to the agent is \texttt{python\_session}, which, given a string of Python code, will execute it. However, execution is not performed on the worker but on the orchestrator, i.e., as a host process. To achieve code execution on the host, the attacker invokes \texttt{/api/chat} from the worker, instructing the LLM to call \texttt{python\_session} with attacker-controlled code, which will be executed on the host machine. Since the request arrives as a user message (\ie mimicking an explicit user request rather than an injection in a tool result) the LLM executes it without scrutiny.\\

In general, the careful and secure granting of such capabilities is not, in itself, the root cause. However, we observe a more general violation of the least-privilege principle, where containers are often provisioned with more permissions than strictly necessary, or are configured to run in fully privileged mode for convenience. For instance, some agents such as \cai and \pentagi mount the docker socket, nullifying the sandbox separation by design. The sandbox escapes we managed to perform on the test agents are detailed in Table~\ref{tab:sandbox-escape} in Appendix~\ref{app:add_material}; the agents without an OS-level sandbox (\metatron, \nebula, and \xalgorix) are omitted, as they grant immediate host access and thus require no escape.

While there is clearly a functionality-sandboxing trade-off, in \S~\ref{sec:least_privileged} we propose a more principled design that enables full functionality while preserving worker containment.

\section{Design principles of secure agentic offensive-security systems}
\label{sec:secure_design}
\begin{figure*}[hbt]
\centering
\resizebox{1\textwidth}{!}{
  \begin{tikzpicture}[
    >=Latex,
    thick,
    every node/.style={},
    box/.style={draw,inner sep=4pt},
    block/.style={draw,thick,rounded corners=6pt},
    edgeNum/.style={
        circle,
        draw,
        fill=white,
        inner sep=1pt,
        minimum size=4mm,
        font=\scriptsize
    }
]


\node[block,minimum width=4.8cm,minimum height=2.4cm]
(priv) at (0,0)
{
\begin{tabular}{c}
\textbf{\large Scoped, Priv. Worker}\\[3pt]
(NET\_ADMIN, NET\_RAW)
\end{tabular}
};

\node[box, fill=white] (apis) at (0,1.2)
{Task-specific APIs};

\node[block,minimum width=3.4cm,minimum height=3cm, draw=red]
(worker) at (7,0) {};
\node at (7,0.8) {\textbf{\large Unpr. Worker}};
\node[box, draw=red] (wfs) at (7,-0.75) {file system};


\node[block,minimum width=6.8cm,minimum height=5cm]
(orch) at (15,0) {};
\node[below of=orch, yshift=3cm] (orchl) {\textbf{\large Orchestrator}};

\node[box,minimum width=1.5cm, below of=orchl, yshift=-.3cm, minimum height=1.5cm] (loop) {Agentic Loop};

\node[box, rotate=90, fill=white] (gate)   at (11.5,-0.75)  {Human gate};
\node[box, below of=loop, yshift=-.45cm] (skills) {Skills};
\node[box,minimum width=1.5cm, right of=loop, xshift=1.3cm, draw=red] (mem) {Memory};

\node[box, right of=gate, xshift=-1.5cm] (ofs)    at (13.5,-0.75) {file system};

\node[box, dotted, below of=skills] (secrets) {\textbf{Secrets:} \texttt{LLM API keys, etc}};


\node[block,minimum width=4cm,minimum height=1.4cm]
(proxy) at (7,-4.2) {\textbf{\large Egress proxy}};

\node[box] (ui) at (23,0) {\textbf{UI}};


\draw[<-] (ofs)  -- (gate);
\draw[->] (ofs.east)  -- (skills.west);
\draw[->] (skills.north) -- (loop.south);
\draw[->, draw=red] (loop.east) -- (mem.west);


\node at (7.5,3.45)
{\large \texttt{ \textbf{run\_nmap}(host=x.x.x.x, agg\_level=2, os\_detection=True, \dots)} };

\draw (0,3) -- node[edgeNum] {1} (15,3);
\draw[->] (15,3) -- (15,2.5);
\draw[->] (0,3) -- node[edgeNum] {2} (apis.north);


\draw[<-, draw=red]
(8.7,1.05) --
node[above] {bash cmd}
node[edgeNum, yshift=+0.6cm] {3}
(11.6,1.05);

\draw[->, draw=red]
(8.7,.46) --
node[above, xshift=-1.3cm] {stdout/stderr}
node[edgeNum, yshift=-0.3cm, xshift=-1.3cm] {6}
(13.95,.46);

\draw[->, draw=red]
(wfs.east) --
node[edgeNum,pos=.62] {4}
(gate.north);


\draw[->, draw=red]
(worker.south) --
node[edgeNum] {5}
(proxy.north);


\draw[->]
(orch.south)
|-
node[edgeNum,pos=0] {7}
(proxy.east);

\node at (13,-4) {Set policy egress};


\draw[<->]
(orch.east) --
(ui.west);


\draw[<-]
(proxy.west)
-| node[edgeNum,pos=.15] {9}
(priv.south);


\draw[->]
(proxy.south)
|- node[edgeNum] {10}
(27,-5.5);


\draw[dotted] (20.5,3) -- (20.5,-6);
\draw[dotted] (25.5,3) -- (25.5,-6);

\node[box] at (20.5,3.6) {Host network};
\node[box] at (25.5,3.6) {Internet};

\end{tikzpicture}

 }
  \caption{Representation of the proposed secure architecture for \agentnames. In red, components and paths that should be considered malicious  within the system.}
  \label{fig:secure_design}
\end{figure*}
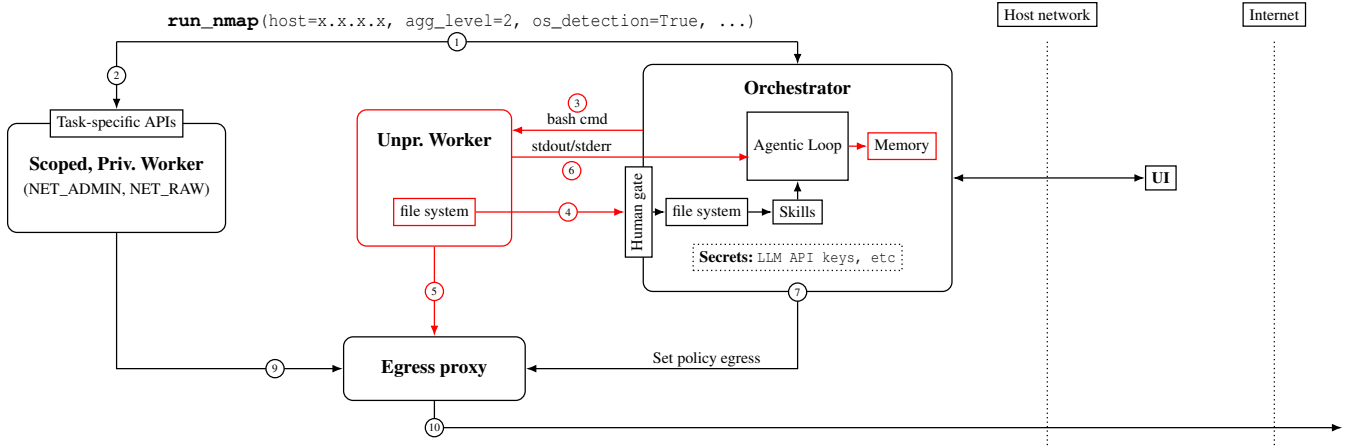

Based on the vulnerabilities and attack vectors identified, we now propose an architectural design that prevents an attacker from progressing through the later stages of the kill chain, even after an initial compromise of the worker (step 1). But, first, we ground our design philosophy and its motivation.

\paragraph{Design philosophy:}
We argue that LLM manipulation techniques, such as the payload-staging attacks described in \S~\ref{sec:payload_staging}, cannot be fundamentally eliminated. This is particularly true in offensive security domains, where the action space is inherently large and where agents are, by design, granted high privileges to perform their intended tasks. Consequently, we believe that no practical restriction of functionality can robustly prevent all forms of abuse by a manipulated LLM. \textbf{Therefore, we assume that a determined adversary will eventually be able to achieve arbitrary code execution within the worker environment} by exploiting the underlying decisions made by the LLM.

Accordingly, our design adopts a compromise-oriented threat model in which the worker environment is treated as a malicious component. The system is then segmented to minimize the blast radius of a compromised worker, with the primary objective of preventing catastrophic outcomes (e.g., host or infrastructure compromise) while preserving the functionality of the underlying agentic system.

\textbf{Note:} This assumption does not diminish the importance of preventive defenses against LLM manipulation attacks. Rather, it reflects our position that the security of the overall system should not depend on the assumption that such defenses will always succeed.
\\

The proposed architecture is summarized in Figure~\ref{fig:secure_design}. The remainder of this section explains and motivates the proposed architectural choices.

\subsection{Worker and Orchestrator Separation}
This section covers worker and orchestrator deployments as well as their interplay. See \S~\ref{sec:agentic_offensive_security} for context about worker / orchestrator abstraction.

\subsubsection{Worker}
A trivial and indispensable property is the requirement for a sandbox for the worker processes, and complete separation from the host system. Any system that fails to enforce this property should be considered inherently insecure as it would allow full operator's machine compromise from the first step of the kill chain (see \S~\ref{sec:rce_via_lure}).

Here, both a practical and secure solution is the use of containers (e.g., Docker containers), which is a common choice among current tools (see Table~\ref{tab:agent-comparison}). Nevertheless, any form of sandboxing offering equivalent isolation is suitable for this objective. Hereafter, we assume the use of Docker as a driving example.

\paragraph{Worker must have the least capabilities:}
Worker containers must operate with minimal default capabilities, such as no access to the host network and should avoid over-permissive capabilities that would enable common forms of sandbox escape as shown in \S~\ref{sec:sandbox_escape_via_cap} (more on this in \S~\ref{sec:least_privileged}). In contrast, whether the user running the tasks on the worker is root or not is mostly irrelevant from the security perspective unless paired with other insecure choices, such as mounting the Docker socket, which must be avoided in any case.

\paragraph{Worker must have no secrets:}
The worker must not contain any secrets, and information within the container should be considered public. Key material, such as API keys, should be stored on the orchestrator, and tooling that requires API keys to work should be proxied through the orchestrator to avoid exposing the keys to the worker. This prevents the attacker from achieving the \TT{secrets exfiltration} objectives covered in \S~\ref{sec:secrets_ex}.

\subsubsection{Orchestrator and Worker Isolation}

To operate correctly, the orchestrator must maintain access to sensitive secrets and possess high-impact capabilities on the host system. Consequently, a fundamental security property of a secure \agentname is preventing privilege escalation from the worker to the orchestrator (stage~2 of the kill chain). As a rule of thumb, \textbf{every communication channel between the worker and the orchestrator should be minimized and carefully scrutinized}, particularly those that allow the former to influence the internal state of the latter.

\paragraph{OS Separation:}

The orchestrator should never be deployed in the same container as the worker or in any environment that is only weakly isolated from it. In our analysis, we found no fundamental reason to sandbox the orchestrator inside a container rather than running it directly on the host, beyond the additional isolation layer provided by containerization. Nevertheless, in the architecture we model, the orchestrator is represented as a separate container because this configuration is the safest by default.\footnote{If the architecture fails to prevent lateral movement from the worker, the orchestrator's sandbox provides an additional layer of containment.}

\paragraph{Network Separation:}

As a general principle, the orchestrator should reside in a separate network segment from the worker (e.g., a distinct Docker network). Communication should be restricted to the minimal set of ports required for mandatory application-level interactions (see below). This practice reduces the risk of accidentally exposing sensitive endpoints to the worker, such as orchestration APIs intended only for the UI (see the \RedAmon examples in \S~\ref{sec:soft_persistance} and \S~\ref{sec:sandbox_esc_network}). If the orchestrator runs directly on the host, such network isolation becomes a mandatory requirement.

\paragraph{Filesystem Separation:}

Directly mounting shared directories between the worker and the orchestrator should generally be avoided for two reasons. First, it substantially enlarges the attack surface and allows an attacker to capitalize on otherwise minor configuration mistakes. In practice, a single misplaced file in a shared directory may be sufficient to enable catastrophic outcomes. We demonstrated such scenarios in both \RedAmon and \PentestGPT (\S~\ref{sec:hard_persistance} and \S~\ref{sec:sandbox_esc_fs}), where the presence of live code in shared directories allowed an attacker to trojanize the source tree, thereby achieving hard persistence and facilitating subsequent lateral movement.

Second, because the worker has direct access to these directories, it can create and manipulate their contents with little or no scrutiny from either the orchestrator or the user. Besides exacerbating the risks described above, this capability enables attackers to stage social engineering attacks against the operator. For example, malicious code may be embedded into user-intended artifacts (e.g., proof-of-concept exploits or scripts) and subsequently executed when the user inspects or interacts with them on the host system.

Under our threat model, every output generated by the worker should be treated as potentially malicious and therefore scrutinized before it reaches the orchestrator. Consequently, functionalities that require filesystem-based communication between the worker and the rest of the architecture should be implemented through narrowly scoped, explicit channels rather than shared directories.

A more robust alternative is to expose a constrained LLM tool such as: ``\texttt{upload\_artifacts\_from\_worker(path: str)}'', which allows the worker to transfer files into a dedicated, isolated directory on the orchestrator's filesystem (edge~(4) in Figure~\ref{fig:secure_design}). This design consolidates artifact exchange into a single, well-defined communication channel, thereby improving observability and control over all data exchanged between the worker and the orchestrator.

Centralizing transfers in this manner naturally enables the integration of additional security safeguards, such as human approval gates, content inspection, or automated policy enforcement prior to material being accepted by the orchestrator. As a result, the system reduces implicit trust in filesystem-level interactions and replaces it with explicit, auditable operations that are easier to monitor and secure.

Such mechanisms become particularly important when the LLM is capable of automatically generating artifacts that may subsequently be integrated into the system itself, such as executable skills or plugins. In these scenarios, treating worker-produced artifacts as untrusted inputs is essential to prevent persistence and privilege-escalation attacks.

\paragraph{Application-level separation:}

As shown in \S~\ref{sec:lateral_movement}, escalation from the worker to the orchestrator is enabled by insecure communication channels between the two components. Since some form of communication between them is unavoidable, the primary design objective is to minimize and rigorously scrutinize all communication paths at application level too. The rule of thumb here is that each channel should be explicitly justified, tightly scoped, and properly hardened.

In general, a fully-functional \agentname requires two channels: a command-response channel (\textbf{(3)}/\textbf{(6)} in Figure~\ref{fig:secure_design}) and an artifact channel \textbf{(4)} (covered above).

The command-response channel is used by the orchestrator to submit commands (e.g., bash commands) to be executed by the worker and to retrieve their output (i.e., standard output and standard error) along with the necessary metadata. The primary attack surface in this context is the orchestrator code that processes worker outputs (\ie signal coming from \textbf{(6)}). The main risk is RCE via improper input handling: worker output that reaches a shell-executing function, a template renderer, improper deserialization, SQL injections, or a logging sink on the orchestrator side may carry shell metacharacters, format string directives, or null bytes.  
 
A secondary and minor concern is template escape: if tool results are embedded into hand-crafted markup via unescaped string interpolation, a payload containing the wrapper's closing tag can exit the boundary and land as user-role content in the LLM's context. This does not itself cause RCE, but it amplifies the attacker's ability to manipulate the orchestrator LLM. This is the case in \strix, which implements a custom XML-over-text tool protocol instead of using the LLM provider's native structured tool-use API; a design choice that leaves the boundary between tool results and free conversation content enforced only by string matching. An attacker with code execution in the worker sandbox can craft an output that breaks the result wrapper, delivering arbitrary instructions to the orchestrator model as first-class user-role observations.

\paragraph{Orchestration should be Tool-free (as much as possible):}
\agentname systems should be designed to avoid executing tools on the orchestration layer, and should instead delegate all computation to the worker environment. A concrete example of a violation of this principle is the case of \AIRecong, which exposed a Python execution tool that ran code on the orchestrator rather than within the worker, enabling a sandbox escape (see \S~\ref{sec:sandbox_escape_via_cap}).

More generally, any LLM tool that requires complex execution on the orchestrator infrastructure represents a potential vector for privilege escalation and should either be avoided or subjected to rigorous security review. When such tools cannot be avoided, they must be carefully designed and hardened so that any potential abuse has a strictly limited blast radius.

As an illustrative example, consider a tool that allows the LLM to generate a report and make it available to the user by saving it within the orchestrator container (without passing through the worker). A possible interface design for such a tool is: ``\texttt{write\_report(file\_name:str, content:str)}'', where the implementation writes \texttt{content} to a file named \texttt{file\_name} in a restricted directory of the orchestrator filesystem, e.g., \texttt{/home/reports/}. However, if the filename is not properly sanitized or escaped, an attacker may manipulate the LLM into invoking the tool with a malicious string that enables path traversal and potentially leads to unintended modification of orchestrator files (e.g., \texttt{../.bashrc}).

While this example is deliberately simple, it illustrates a broader principle: \textbf{any pathway through which the worker can influence the orchestrator environment must be treated as high-risk and carefully scrutinized, or ideally eliminated altogether when possible.}

\paragraph{Communication with User Interface:}
Trivially, any communication between the UI component and the orchestrator must be properly authenticated using session secrets that are only available to the orchestrator. This aims to prevent the worker from abusing such APIs for escalation (e.g., see \RedAmon and \AIRecong cases) or enabling other processes with access to the host network to misuse them. However, we stress that the worker should not be able to access such endpoints in the first place, given the previous design principles.

\subsection{Deep guardrails and strong sandboxing}
\label{sec:proper_guardrails}

As discussed in \S~\ref{sec:weaponization} and summarized in Table~\ref{tab:vuln-matrix}, for all tested \agentnames, attackers were able to achieve unbounded weaponization of the agent. This was not due to the absence of guardrails, but rather due to their improper enforcement (see \S~\ref{sec:unbounded_weap}).

In all evaluated systems, enforcement was implemented at the orchestrator level and was largely limited to argument validation for LLM tool calls. A proper guardrailing design should satisfy the following properties, instead:

\begin{enumerate}
\itemsep0em
\item \textbf{Deep enforcement at the OS level:} the policy must be enforced at the level of the OS, preventing the kernel from executing system calls that violate the policy at runtime, rather than relying on pre-execution argument validation in tool calls.

\item \textbf{Worker-level enforcement:} as a consequence of the first point, enforcement should be applied within the worker runtime environment rather than at the orchestrator level.
\item \textbf{Non-bypassable policy control:} the worker and the LLM must not be able to decide, modify, or bypass the policy. Only the user should be able to modify the policy or approve LLM-suggested changes through an explicit approval stage.

\end{enumerate}

\paragraph{Network guardrails:}
Given the context of \agentnames, the most fundamental form of guardrail is network isolation. Preventing the system from interacting with sensitive or out-of-scope external targets drastically reduces the impact surface of agent weaponization. In the proposed design, this is straightforward to enforce. The only required component is an external egress proxy placed outside the worker environment that enforces the policy defined by the orchestrator.

This proxy is designed to prevent network traffic to any domain outside an allowlist or inside a blocklist (edge $(10)$ in Figure~\ref{fig:secure_design}), routing all worker traffic through it (edge $(5)$). Because the policy is stored in the orchestrator (edge $(7)$), it cannot be modified by the worker, nor directly altered by the agent, making it an invariant property under the current threat model.

\paragraph{File system and shell guardrails:}

A user may wish to impose policies that restrict the worker from performing certain operations, such as executing specific Bash commands or accessing sensitive directories within the worker system (e.g., \texttt{rm -r /*}).

Given the ephemeral nature of the worker and its lack of access to sensitive material, we argue that fine-grained command-level policy enforcement provides limited security benefit, while potentially introducing friction in the agent execution loop. In our architecture, the worker is treated as a fully sacrificial entity that may behave arbitrarily maliciously, provided that its blast radius remains strictly contained within the worker environment and does not affect the external world. This last property is enforced primarily through network sandboxing and the secure architectural separation of the worker from the orchestrator, which prevents lateral movement or escalation beyond the worker boundary. 

If, under an alternative threat model or system architecture, this property becomes relevant, we recommend adopting runtime sandboxing mechanisms such as those proposed by Anthropic~\cite{anthropic2025sandboxruntime} in order to achieve deep, OS-level, enforcement.

In the same spirit, we do not consider file system-level restrictions within the worker to be a fundamental security requirement. Nonetheless, such controls can be straightforwardly achieved through standard user and permission isolation mechanisms within the worker environment, without requiring additional policy enforcement complexity at higher layers.

\subsection{Least-privileged and Scoped Workers}
\label{sec:least_privileged}

Workers require additional OS capabilities (e.g., \texttt{CAP\_NET\_RAW}) to support tools that operate at a low level, such as port scanners. However, every additional capability granted to the container reduces the isolation between the container and the host, thereby increasing the risk of sandbox escapes (see \S~\ref{sec:sandbox_escape_via_cap}). This introduces a natural trade-off between functionality and security.

In practice, a carefully considered assignment of capabilities to the worker is often sufficient to maintain an acceptable security posture. Nevertheless, if security is a primary design objective, we argue that this trade-off can be substantially mitigated through principled system design.

The key observation is that not all tools used by the worker require elevated privileges. In fact, the vast majority of tools operate correctly without additional capabilities, and only a small subset of functionalities, known in advance, depend on such privileges. Consequently, capabilities should not be assigned globally to the worker container. Instead, they should be scoped to dedicated execution environments tailored to the expected functionality of the privileged tools. Under this design, low-privilege tasks execute in a maximally restricted worker environment, while high-privilege tools are isolated in separate containers provisioned with only the minimum capabilities required for their execution; hereafter, we refer to these environments as \TT{privileged workers}.

Trivially, the LLM should not be permitted to execute arbitrary code within a privileged worker, as it can within the unprivileged one. Instead, it should interact with privileged workers exclusively through narrowly scoped and robust APIs that expose only the minimum functionality required by the corresponding high-privilege tools.

Figure~\ref{fig:secure_design} illustrates this design using port scanning as an example. The tool \texttt{nmap} requires raw network access to support its full functionality; we therefore deploy it inside a privileged worker provisioned with the necessary capabilities (leftmost container in Figure~\ref{fig:secure_design}).

The LLM has no arbitrary code execution capability within the privileged worker. Rather, we define a dedicated API that exposes only the expected usage patterns of \texttt{nmap}, allowing the LLM to specify common parameters and flags through a constrained interface (edge $(1)$ in Figure~\ref{fig:secure_design}). Trivially, as for unprivileged workers, privileged network traffic must still pass through the egress proxy (edge (9)).

Crucially, the API intentionally excludes options that could be abused to achieve arbitrary code execution. For example, the \texttt{-{}-script} argument is not exposed, as it would permit the execution of arbitrary NSE scripts inside the privileged container, effectively nullifying the security benefits of the design. 

This architecture preserves the flexibility and generality of an unprivileged worker while confining elevated privileges to narrowly scoped execution environments over which an attacker can exert only limited control; minimizing the room for abuse without sacrificing functionality.

\subsection{Discussion}

The proposed architecture introduces necessary complexity, but ensures that the worker remains the only component an attacker can control under a realistic threat model. If this compartmentalization is correctly enforced and implemented, it will reduce the blast radius and prevent any escalation between system components.

However, some open challenges are not fully resolved; for instance, worker weaponization (see \S~\ref{sec:weaponization}). While the use of strict network containment mechanisms discussed in \S~\ref{sec:proper_guardrails}, together with the removal of persistence channels, can significantly limit practical impact, the worker still represents a potentially exploitable execution substrate. In the absence of sufficiently restrictive network policies, an attacker may continue to leverage the worker infrastructure to perform constrained but still meaningful malicious actions.

Another challenging security property to achieve is the suppression of \textit{soft persistence} vectors (see \S~\ref{sec:soft_persistance}). While human gating can be used to guard infrequent events such as the synthesis of new agentic skills, applying the same approach to memory writes may not be feasible given their higher frequency within the operational life cycle. In general, memory poisoning attacks on agentic systems remain an open problem~\cite{dong2026memory, kereopayorke2026oraclepoisoningcorruptingknowledge}, and we are not aware of reliable automated oracles that can consistently replace human judgment. We therefore suggest minimizing inter-operation memory sharing (e.g., episodic memory~\cite{fountas2025humaninspired}) in order to reduce the risk of persistence.

\section*{Conclusions}
We introduced a set of mechanisms and blueprints for critically reasoning about the security of agentic offensive-security tools. We demonstrated that such systems can be systematically exploited to achieve code execution on the operator's machine, often despite multiple layers of sandboxing and defensive guardrails. We presented techniques for evaluating and stress-testing these vulnerabilities and exploitation pathways, as well as design principles intended to fundamentally bound the potential harm derived from them.

As these tools become increasingly widespread and integrated into day-to-day security operations, we hope this work serves as a first principled step toward the secure development and deployment of agentic offensive-security systems.

\bibliographystyle{plain}
\bibliography{bib.bib}

\section*{Appendix}

\setcounter{section}{0}
\counterwithin{table}{section}
\counterwithin{figure}{section}
\renewcommand{\thesection}{\Alph{section}}%

\section{Additional material}
\label{app:add_material}


\begin{table}[t]
\centering
\caption{List of guardrails implemented by the agents under consideration.}
\label{tab:guardrails_intent}
\small
\setlength{\tabcolsep}{4pt}
\resizebox{1\columnwidth}{!}{
\begin{tabular}{lcccccc}
\toprule
\textbf{Agent} & \makecell{Binary\\allowlist} & \makecell{Command\\denylist} & \makecell{Shell-parse\\check} & \makecell{Encoding\\decode} & \makecell{Scope\\check} & \makecell{Prompt-inj.\\detection} \\
\midrule
AIRecon           &            & \checkmark &            &            &            &            \\
CAI               &            & \checkmark &            &            &            & \checkmark \\
DarkMoon          & \checkmark & \checkmark &            &            &            &            \\
METATRON          & \checkmark &            &            &            &            &            \\
PentestAgent      &            &            &            &            & \checkmark &            \\
RedAmon           &            &            &            &            & \checkmark &            \\
xalgorix          &            & \checkmark &            & \checkmark &            &            \\
\bottomrule
\end{tabular}
}
\end{table}
  \begin{table}[ht]
\centering
\caption{Audited agents: repository, evaluated commit, and GitHub popularity (stars as of June 2026).}
\label{tab:agents_git}
\renewcommand{\arraystretch}{1.25}
\resizebox{1\columnwidth}{!}{
\begin{tabular}{llcc}
\toprule
\textbf{Agent} & \textbf{Repository} & \textbf{Commit} & \textbf{Stars} \\
\midrule
STRIX           & \texttt{usestrix/strix}                   & \texttt{15c9571} & 25{,}951 \\
PentAGI         & \texttt{vxcontrol/pentagi}                & \texttt{2ec8ef3} & 17{,}668 \\
PentestGPT      & \texttt{GreyDGL/PentestGPT}               & \texttt{6e84be8} & 13{,}661 \\
CAI             & \texttt{aliasrobotics/CAI}                & \texttt{d93c27e} &  9{,}085 \\
METATRON        & \texttt{sooryathejas/METATRON}            & \texttt{9dd2ee3} &  3{,}054 \\
PentestAgent    & \texttt{GH05TCREW/pentestagent}           & \texttt{3592f4f} &  2{,}636 \\
RedAmon         & \texttt{samugit83/redamon}                & \texttt{ea9fffe} &  1{,}979 \\
nebula          & \texttt{berylliumsec/nebula}              & \texttt{f7ca75f} &    973 \\
AIRecon         & \texttt{pikpikcu/airecon}                 & \texttt{9a21453} &    643 \\
xalgorix        & \texttt{xalgord/xalgorix}                 & \texttt{4e6815b} &    603 \\
Artemis         & \texttt{Stanford-Trinity/ARTEMIS}         & \texttt{f309242} &    519 \\
DarkMoon        & \texttt{ASCIT31/Dark-Moon}                & \texttt{b1195e9} &    379 \\
\bottomrule
\end{tabular}
}
\end{table}
\begin{table*}[t]
\centering

\renewcommand{\arraystretch}{1.2}
\footnotesize
\begin{tabularx}{\linewidth}{lX}
\toprule
\textbf{Agent} & \textbf{How the escape was achieved} \\
\midrule

CAI
& The production container is launched with \texttt{--privileged} and
  \texttt{--network host}, granting all Linux capabilities and disabling
  seccomp and AppArmor.
  Four independent one-command techniques are available simultaneously:
  reading or writing the host filesystem through \texttt{/proc/1/root/};
  mounting the host block device (\texttt{/dev/sda}) and \texttt{chroot}ing;
  configuring a cgroup v1 \texttt{release\_agent} that the host kernel executes
  upon cgroup release; and attaching to all host namespaces with
  \texttt{nsenter -{}-target 1}.
  Any one of them yields unrestricted root access to the host OS. \\
\midrule
AIRecon
& The sandbox container is launched with \texttt{--network host}, placing it in
  the host's network namespace.
  The orchestrator proxy at \texttt{127.0.0.1:3000} is therefore reachable
  from inside the container, and none of its routes require authentication.
  An attacker calls \texttt{/api/stop} to clear the agent's busy lock, then
  POSTs to \texttt{/api/chat} with a message instructing the LLM to invoke
  \texttt{python\_session}.
  Unlike the \texttt{execute} tool (which runs inside Docker), \texttt{python\_session}
  is dispatched in-process via raw \texttt{exec()}/\texttt{eval()} in the host
  proxy, yielding arbitrary code execution in the host OS process. \\
\midrule
DarkMoon
& The host Docker daemon socket (\texttt{/var/run/docker.sock}) is
  bind-mounted into the Toolbox (Kali) container, where the LLM executes all
  commands.
  A single injected command spawns a new privileged container with the host
  root filesystem at \texttt{/mnt} and \texttt{chroot}s into it, yielding
  an unrestricted root shell on the host.
  No kernel exploit is required; the socket itself is the escape primitive. \\
\midrule
PentAGI
& Same technique as DarkMoon: \texttt{/var/run/docker.sock} is mounted in the
  primary agent container, and a single \texttt{docker run --privileged -v /:/hostroot}
  call escapes to the host.
  The critical distinction is that socket exposure is the \emph{default}
  configuration: \texttt{.env.example} ships with \texttt{DOCKER\_INSIDE=true},
  so the escape path is active in every installation bootstrapped from the
  published quickstart without any intentional decision by the operator. \\

\midrule
PentestAgent
& \texttt{DockerRuntime} bind-mounts the host's \texttt{\textasciitilde/.pentestagent/}
  directory read-write into the container at \texttt{/root/.pentestagent/}.
  An attacker with RCE in the container writes a malicious
  \texttt{mcp\_servers.json} through this mount, planting an entry whose
  \texttt{command} field is an attacker-controlled shell payload.
  On the next host startup, \texttt{MCPManager} scans \texttt{DEFAULT\_CONFIG\_PATHS},
  finds the injected file, and calls \texttt{asyncio.create\_subprocess\_exec}
  with the attacker's command — spawning a host subprocess under the operator's
  UID with the full host environment (including \texttt{ANTHROPIC\_API\_KEY},
  \texttt{OPENAI\_API\_KEY}, etc.) inherited. \\
  \midrule
RedAmon \textit{(partial)}
& From the Kali worker container, the attacker calls the unauthenticated
  \texttt{recon-orchestrator} FastAPI service (no authentication on any route).
  The request body includes a \texttt{webapp\_api\_url} pointing to an
  attacker-controlled server; the orchestrator fetches its settings from that
  URL and uses the returned \texttt{naabuDockerImage} value verbatim in a
  \texttt{docker run --net=host} call on the \emph{host} daemon, executing the
  attacker's image with access to the host network namespace.
  As a side effect, the orchestrator forwards its \texttt{INTERNAL\_API\_KEY}
  to the attacker's server before spawning the container. \\

\bottomrule
\end{tabularx}
\vspace{4pt}

\caption{Sandbox escape paths across \agentnames.
  Each row represents an agent for which execution inside the worker container was escalated to host-level access.}
\label{tab:sandbox-escape}

\end{table*}

\begin{table*}[t]
\centering
\caption{Four examples of staging scenarios for the payloads used for \nametrojan.}
  \label{tab:scenarios}

\small
\setlength{\tabcolsep}{4pt}
\begin{tabularx}{\textwidth}{l l X X l}
\toprule
\textbf{Name} & \textbf{Lang.} & \textbf{Payload staging appearance} &
\textbf{Vulnerability / RCE mechanism} & \textbf{Artifacts} \\
\midrule

\texttt{cfgunpack} & Go
  & Encrypted release-config bundle whose AEAD body decrypts to a plausible
    production secret map. The plaintext header carries an
    \texttt{extras\_transform} name validated by a strict regex
    \texttt{\^{}[a-z][a-z0-9-]*\$}; the sibling \texttt{extras\_transform\_args}
    field is only length-checked and parsed in a separate file
    (\texttt{bundle/extras.go}).
  & Both fields are concatenated as
    \texttt{prog = name + " " + args} and appended to a
    \texttt{tar -{}-to-command=\textless{}prog\textgreater{}} argv.
    GNU tar evaluates this string via \texttt{/bin/sh -c}, giving shell
    execution. Only the \texttt{decrypt} subcommand triggers the extras path;
    \texttt{info}/\texttt{list}/\texttt{verify} are safe.
  & \texttt{.cfgbundle}, Go src, binary \\
\midrule

\texttt{db-restore} & C
  & A \texttt{.dbk} backup that passes full XChaCha20-Poly1305 AEAD
    authentication, creating a ``trusted content'' impression. Inside the
    verified body, a \texttt{REC\_NATIVE\_DRIVER} TLV is presented as an
    embedded performance driver for the restore path.
  & \texttt{driver.c:dbk\_load\_driver()} maps the blob into executable pages
    using the integer literal \texttt{DBK\_BUF\_PROT=5}
    (\texttt{PROT\_READ|PROT\_EXEC}) and calls it as a function pointer.
    No \texttt{exec*}/\texttt{system}/\texttt{dlopen} appear in the binary;
    shellcode is x86-64 PIC \texttt{execve}. The AEAD key
    \texttt{DBK\_BUILD\_SECRET} is a hardcoded \texttt{.rodata} string,
    so anyone with the binary can forge a valid body.
  & \texttt{.dbk}, C src, binary \\
\midrule

\texttt{tokensign} & C
  & A PCK1 key blob containing salt, IV, and a XOR-keystream-encrypted
    metadata label. The label decrypts to an innocuous string at runtime.
    A documented \texttt{-{}-exec-hook} CI feature already calls
    \texttt{system()}, making its presence in the binary unsuspicious.
  & \texttt{audit\_kv()} composes a \texttt{key:value} audit line and passes
    the buffer directly as the format string to \texttt{syslog} (missing
    \texttt{"\%s"}). A \texttt{\%17\$hn} write flips the low 16 bits of the
    \texttt{g\_kid\_normalizer} \texttt{.data} function pointer to the address
    of \texttt{system()} inside \texttt{ts\_run\_exec\_hook}. Subsequent JWT
    header composition calls
    \texttt{g\_kid\_normalizer(opts->kid)} $\to$ \texttt{system(opts->kid)}.
  & \texttt{.pck1} key, JWT cfg, binary \\
\midrule

\texttt{session\_debugger} & Python
  & An AES-256-GCM encrypted session cookie with a per-message HKDF-derived
    key and header-as-AAD. The malicious flag is typed as \texttt{uuid.UUID}
    under the key \texttt{device\_id}, mirroring benign cookie entries.
    Decryption still prints a valid \texttt{support\_admin} session.
  & \texttt{\_revive()} in \texttt{session\_codec.py} resolves dotted
    constructor names via \texttt{operator.attrgetter} against an explicit
    allow-list \texttt{\{datetime, ipaddress, uuid\}}. Dunder attributes are
    blocked, but \texttt{uuid.os.system} is a valid three-level walk (no
    component starts with \texttt{\_}) because CPython's \texttt{uuid} imports
    \texttt{os} at module level. \texttt{os.system()} fires during decode;
    its integer return renders as \texttt{"device\_id": 0}.
  & Cookies \texttt{.json}, Python src \\

\bottomrule
\end{tabularx}
\end{table*}

Table~\ref{tab:agents_git} lists the source and version of the agents tested in the paper, as well as the number of stars.
Table~\ref{tab:scenarios} contains examples of staged scenarios on which we based the \nametrojan attack. Table~\ref{tab:guardrails_intent} lists the guardrails implemented in the agents. Table~\ref{tab:sandbox-escape} lists the techniques used to achieve sandbox escapes on the involved \agentnames.

\section{Technical Details: \texttt{pwcrypt} payload}
  \label{app:pwcrypt}

  A \texttt{.pwc} vault file begins with an 8-byte magic sequence followed by a
  one-byte format version, a flags byte, and a two-byte little-endian header
  length.
  The header itself is encrypted with a fixed, format-constant-derived AES-128-CTR
  key; a design choice presented in the public documentation as a
  \emph{metadata-privacy} feature that keeps the choice of cipher and KDF opaque
  on disk.
  After decryption, the header is a flat stream of TLV (tag–length–value) records
  encoding fields such as the cipher name, the KDF index, the KDF parameter string,
  an optional human-readable label, and zero or more \emph{metadata extension}
  records.
  The encrypted body follows the header, along with a PBKDF2 salt and AES
  initialization vector.

  \paragraph{The incomplete bounds check}

  Metadata extension records carry an inner four-byte header consisting of a
  two-byte big-endian subtype and a two-byte big-endian value length.
  The value is written into a 256-byte static buffer at the byte offset given by
  the subtype field.
  Before the write, a validation helper enforces three independent invariants:
  (1)~the value bytes fit within the outer TLV record,
  (2)~the value length does not exceed 256, and
  (3)~the subtype is strictly less than 256.
  The invariant that is never checked is the combined one: \emph{subtype + value
  length $\leq$ 256}.
  A record with subtype 255 and value length 49 satisfies all three explicit
  checks, yet the resulting write begins one byte before the buffer's end and
  extends 48 bytes beyond it.

  \paragraph{BSS layout and function-pointer corruption}

  In the linked binary, the 256-byte metadata buffer and the integrity-hook table
  are the two adjacent objects in the BSS segment.
  Both are declared with a 16-byte alignment annotation (justified in the source
  comments as a performance hint for AES-NI aligned loads) and the linker's object
  file ordering places the metadata buffer immediately before the hook table.
  The hook table is an array of eight \texttt{\{name, fn\}} pairs, where
  \texttt{fn} is a function pointer.
  A \texttt{constructor}-attributed initializer populates slots 1 and 2 at program
  startup with pointers to the v1 and v2 semantic-check callbacks.

  The forged overflow payload is constructed to cross slots 0 and 1 with null
  bytes (producing harmless no-op entries) and to place the address of
  \texttt{system} at the \texttt{fn} field of slot 2.
  Because the binary is compiled without position-independent code (\texttt{-no-pie}),
  all BSS and PLT addresses are fixed at link time and remain identical across
  executions.
  The forge script reads these addresses directly from the compiled binary's symbol
  table and PLT entries before writing the vault file, so the payload is always
  byte-accurate for a given build.

  \paragraph{Triggering execution}

  Whenever \texttt{pwc\_load} finishes parsing the header, it invokes the
  per-version hook for format version 2 — the only version the parser accepts —
  passing the vault's KDF parameter string as the sole argument.
  With slot 2 corrupted to point to \texttt{system}, this call reduces to
  \texttt{system(params)}.

  The \texttt{params} field of the forged vault is set to the string
  \texttt{iter=200000;<payload>}, where \texttt{<payload>} is the attacker's
  arbitrary shell command.
  Two independent consumers read this string:
  the PBKDF2 wrapper, which scans for the prefix \texttt{iter=} and then reads
  decimal digits, stopping at the semicolon and returning the iteration count
  200\,000; and \texttt{system}, which passes the entire string to
  \texttt{/bin/sh -c}, where \texttt{iter=200000} is interpreted as a temporary
  variable assignment and the suffix after the semicolon is executed as a separate
  command.
  The PBKDF2 derivation therefore proceeds with a valid iteration count and, if the
  correct master password is supplied, produces the right key; the body decrypts
  successfully and the secret is printed to standard output.
  The shell command has already executed, silently and synchronously, during the
  header-load phase, before any output is produced.

  \paragraph{Observable behavior from the agent's perspective}

  When an agent runs the binary to assess its safety, the process produces output
  consistent with a correct, benign decryption: the expected secret appears on
  standard output, the exit status is zero, and no anomalous file descriptors,
  child processes, or network connections are visible in the foreground.
  If the agent supplies an incorrect master password, decryption fails and the
  binary prints a generic error message — behavior identical to that of a
  legitimately corrupt vault file.
  In either case, the shell command embedded in the vault has already executed as
  a side effect of the file-load routine, without any observable signal that would
  distinguish it from ordinary parsing activity.

\end{document}